\documentclass[12pt]{iopart}

\usepackage{iopams}
\usepackage{float}
\usepackage{amsfonts}
\usepackage{amssymb}
\usepackage{bm}
\usepackage{color}
\usepackage{subfigure}
\usepackage{graphicx}
\usepackage{amsthm}
\usepackage{enumerate}
\usepackage{mathrsfs}
\usepackage[utf8x]{inputenc}
\usepackage{lmodern,textcomp}
\usepackage{amsthm} 
\newcommand{\Da}{D_{\rm a}}
\newcommand{\dW}{\langle\dot W\rangle}

\begin{document}

\title{Energetics of critical oscillators in active bacterial  baths} 


\author{Ashwin Gopal$^{1,2,3}$, \'Edgar Rold\'an$^2$, and  Stefano Ruffo$^{3,4}$}
\address{$^1$ Department of Physics, Indian Institute of Science Education and Research, Pune Dr. Homi Bhabha Road, Pashan, Pune 411008, India}
\address{$^2$ ICTP - The Abdus Salam International Centre for Theoretical Physics, Strada Costiera 11, 34151, Trieste, Italy}
\address{$^3$ SISSA, Via Bonomea 265, 34136 Trieste, Italy}
\address{$^4$ INFN Trieste and ISC-CNR}

\date{\today}

\begin{abstract}
We investigate the nonequilibrium energetics  near a critical point of a  non-linear  driven oscillator immersed in an active bacterial bath. At the critical point, we reveal a scaling exponent of the average power  $\langle\dot{W}\rangle\sim (D_{\rm a}/\tau)^{1/4}$ where $D_{\rm a}$ is the effective diffusivity and  $\tau$ the correlation time of the bacterial bath described by a Gaussian colored noise. Other features  that we investigate are the average stationary power and the variance of the work both below and above the saddle-node bifurcation.  Above the bifurcation,  the average power  attains an optimal, minimum value for  finite $\tau$ that is below its zero-temperature limit. Furthermore, we reveal a finite-time uncertainty relation for active matter  
which leads to values of the  Fano factor of the work that can be below $2k_{\rm B}T_{\rm eff}$, with $T_{\rm eff}$ the effective temperature of the oscillator in the bacterial bath. We  analyze different Markovian approximations to describe the nonequilibrium stationary state of the system. Finally, we  illustrate our  results in the experimental context by considering the example of  driven colloidal particles    in  periodic  optical potentials  within  a \textit{E. Coli} bacterial bath. 
\end{abstract}


\maketitle

\section{Introduction}

Active  systems  are inherently out of equilibrium as their constituent units exchange continuously  energy with their environment (e.g. free energy from ATP consumption) to produce mechanical work (e.g. cellular motion). As a consequence, active matter  leads to a set of  physical phenomena that cannot be described using equilibrium statistical mechanics~\cite{ramaswamy2010mechanics,gompper20202020}, such as collective motion~\cite{julicher1997modeling,kumar2014flocking}, motility-induced phase separation~\cite{cates2012diffusive,narayan2007long,cates2015motility}, 
  noise rectification~\cite{tailleur2009sedimentation,di2010bacterial,galajda2007wall}, etc.
Recent studies within the field of stochastic thermodynamics have investigated the  signatures of active matter in the fluctuations and/or response of passive probes. Prominent examples include estimating the heat dissipation of biological systems from the fluctuations of few nonequilibrium degrees of freedom~\cite{roldan2010estimating,fodor2016far,roldan2018arrow,Caprini_2019},  exploring fundamental bounds for the precision and cost of biomolecular processes \cite{barato2015thermodynamic,gingrich2017fundamental}, and understanding the entropic and frenetic aspects of non-linear response~\cite{maes2020frenesy,basu2015nonequilibrium}.  

Besides its fundamental interest, active matter has recently found prominent applications in the experimental context~\cite{krishnamurthy2016micrometre,bechinger2016active,argun2016non}.  For example, a paradigmatic yet intriguing example is given by the so-called {\em bacterial ratchets}~\cite{di2010bacterial,vizsnyiczai2017light}. These are composed by a microscopic ratchet wheel with asymmetric teeth (i.e.  with broken spatial symmetry) that displays an autonomous rotation when the wheel is immersed in a bacterial bath. 
 Similar  dynamical features like spontaneous  oscillations have been reported in biophysical processes like intracellular transport~\cite{reimann2002brownian} and sound transduction by mechanosensory  hair bundles~\cite{nadrowski2004active,shlomovitz2014phase}, but also in a plethora of artificial nanosystems  such as  phase-locked loops~\cite{gupta1975phase}, and  superconducting Josephson junctions~\cite{barone1982physics,augello2009lifetime,golubev2010statistics}. All this motivates us to study the effects of non-equilibrium fluctuations arising from the interaction of a mesoscopic system with an active bath, in particular at the onset of spontaneous oscillations as is the case of oscillators near a critical point \cite{martin2003spontaneous,nadrowski2004active}.  
 Recent work in stochastic thermodynamics have revamped the interest on the entropy production near phase transitions~\cite{proesmans2020phase,nguyen2020exponential,vroylandt2020efficiency,rana2020precision,seara2019dissipative,martynec2020entropy}, however little is known yet about the non-Markovian effect on the thermodynamics of driven systems embedded in non-equilibrium baths~\cite{speck2016stochastic,horowitz2016work,pietzonka2017entropy,PhysRevX.9.021009,loos2019non}. 
 
\begin{figure}[H]
    \includegraphics[width=\textwidth]{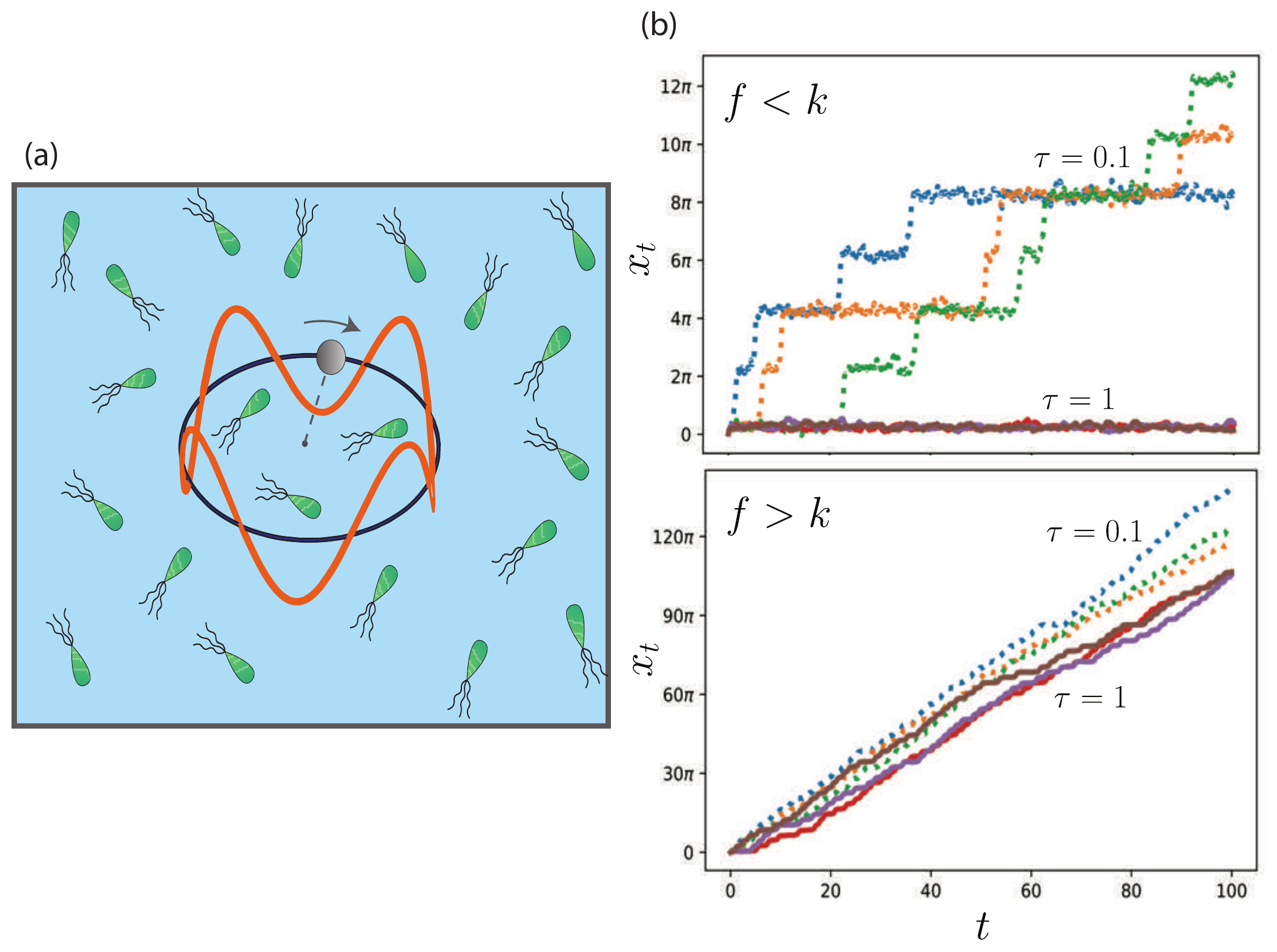}
    \caption{(a) Sketch of the active-matter model studied in this work: a driven non-linear stochastic oscillator in an active bath described by Eqs.~(\ref{eq:1}-\ref{eq:2}). A  Brownian particle (gray sphere)  driven by a torque $f$ is confined in a ring (blue line) and moves in  a non-linear potential (orange line) within    an active bath, here illustrated with green bacteria.  (b) Stochastic trajectories of the total net distance $x_t$ covered by the particle  as a function of time $t$ for bath correlation times $\tau=0.1$ (blue, orange, and green dotted lines)  and $\tau=1$ (red, purple, and violet solid lines): Trajectories below the  critical point  (top, $f=4$) and above the  critical point (bottom, $f=7$) of the oscillator.  The data were obtained from  simulations of Eq.~(\ref{eq:1}-\ref{eq:2}) with time step $\Delta t= 10^{-2}$ using Heun's numerical integration scheme. The rest of the parameters of the simulations were set to  $\gamma=1, L=2\pi, k=6$ and $\Da =1$.}
    \label{Fig.1}
\end{figure}

In this work, we   focus our efforts on the fluctuations of a driven non-linear (so-called {\em Adler}) oscillator in the presence of an active bath. The model describes the motion of a driven Brownian particle in a non-linear periodic potential in the presence of an active bath, see Fig.\ref{Fig.1}(a) for an illustration. The  motion of the particle is governed by the following one-dimensional overdamped Langevin equation
\begin{equation}
   \gamma \dot{x_t} =  f- k \sin{\left(\frac{2\pi x_t}{ L}\right)}+ \eta_t.
\label{eq:1}
\end{equation}
In Eq.~(\ref{eq:1}), $x_t$ denotes the net distance travelled by  the oscillator, $L$ the period of the nonlinear potential, $\gamma$  the friction coefficient, $f$  the external driving torque, $k$ is the strength of the periodic potential, and $\eta_t$ is a Gaussian colored noise with zero mean $\langle\eta_t\rangle=0$ and autocorrelation
\begin{equation}
\langle \eta_t\eta_{t'}\rangle=\frac{\gamma^2\Da}{\tau} \exp(-|t-t'|/\tau),
\label{eq:2}
\end{equation}
i.e. $\eta_t$ is an Ornstein-Uhlenbeck process. 
We remark that the variable $x_t$ is obtained by unwrapping  the phase of the oscillator  $ \theta_t=2\pi x_t/L$ to the real line.  Hence, the dynamics of $x_t$ is akin to the motion of a particle in a one-dimensional tilted periodic potential. 
The choice of the correlation function~(\ref{eq:2}) is far from arbitrary; it is inspired in experimental  work where the motion of a colloid in a bath of \textit{E. Coli} bacteria was described with an instantaneous friction kernel and exponential auto-correlation function~\cite{maggi2017memory,wu2000particle}.

A key feature of the model given by Eqs.~(\ref{eq:1}-\ref{eq:2})   is that, irrespective of the noise properties, the dynamics 
 evolves into a non-equilibrium steady state with time-independent and  homogeneous probability current. The thermodynamic signature of such driven motion is characterized by work dissipation and entropy production.  Little is known about  the stochastic thermodynamics   of nonequilibrium non-linear systems immersed in active baths, for example what is the effect of the  correlation time of the bath  on the average power dissipated and on the work fluctuations?
To  capture the stochastic thermodynamics induced by the active bath, we will also assume the effect of a thermal bath to be negligible compared to the activity of the  bath, which can be experimentally obtained controlling the motility of the bacteria~\cite{maggi2014generalized,argun2016non}. Such approximation is valid    at high concentration of oxygen sources in \textit{E. Coli} bacterial baths~\cite{maggi2014generalized}.

In this article, we  discuss the stochastic thermodynamics of the driven oscillator in a bacterial bath described by Eqs.~(\ref{eq:1}-\ref{eq:2}). In Sec.~\ref{sec:2}, we study the average power inputted into the system in the steady state in the presence of an active bath and compare it with results obtained in the thermal bath limit of small correlation times. In Sec.~\ref{sec:3} we tackle fluctuations and uncertainty of the work done on the system by looking at its variance and Fano factor, and discuss our results in the context of thermodynamic uncertainty relations. In Sec.~\ref{sec:4} we provide  numerical estimates of the power exerted in an  experimentally realizable setup of a colloidal particle immersed in a bacterial bath.   Finally, in Sec.~5  we  review the results and explore  generalizations  to a broader class of stochastic models.


\section{Steady-state average power: critical point and beyond}
\label{sec:2}
 Statistical properties of  the driven oscillator~(\ref{eq:1}) have  been thoroughly studied for the case of $\eta$ given by a  Gaussian white noise~\cite{risken1996fokker}. In the presence of a bath, the variable $x$ can access the entire phase space and fluctuations blur the  separation between  stalling at the fixed point and the oscillating dynamics. As a consequence the average power becomes an analytic function of external torque even at the  critical point, as we will show below. We also  discuss   how the active bath influences on the  power exerted into the system.


The Langevin Eq.~(\ref{eq:1}) yields a  non-Markovian dynamics  for $x$ when $\tau>0$.   For analytical ease it is useful to map the dynamics  to a Markovian system, by adding an additional degree of freedom~\cite{koumakis2014directed,caprini2019active}. Here we will use the Ornstein-Uhlenbeck process  to capture the exponential correlations of the active bath given by Eq.~(\ref{eq:2}).  
The corresponding   two-dimensional Markovian Langevin equation is given by
\begin{eqnarray}
\dot{  x_t} &= &v_f- v_k\sin(2\pi   x_t/ L)+ \eta_t, \label{eom1}\\
    \tau \dot{\eta}_t&= &-\eta_t+\sqrt{2\Da}\xi_t,
    \label{eom}
\end{eqnarray}
where we have introduced the drift variables
\begin{equation}
v_f= f/\gamma,\quad v_k = k/\gamma,
\end{equation}
and $\xi_t$ is a Gaussian white noise with zero mean $\langle\xi_t\rangle=0$ and autocorrelation $\langle\xi_t \xi_{t'}\rangle= \delta(t-t')$.  Following Ref.~\cite{bonilla2019active} we rewrite Eqs.~(\ref{eom1}-\ref{eom}) as an underdamped Langevin equation
\begin{equation}
\hspace{-1cm}
   \gamma\tau\ddot{x}_t= -\gamma \left[ 1+ \frac{2\pi v_k\tau }{L}\cos(2\pi x_t/L) \right]\dot{x}_t +f - k \sin(2\pi x_t/L) + \sqrt{2\gamma k_{\rm B} T_{\rm eff}}\xi_t,
    \label{eum}
\end{equation}
where we have introduced an effective temperature defined as
\begin{equation}
T_{\rm{eff}} = \gamma D_{\rm a}/k_{\rm B}.
\label{eq:Teff}
\end{equation}
Equation~(\ref{eum}) is particularly illuminating to understand the dynamics of the model. It reveals that $m_{\rm eff}=\gamma \tau$ can be interpreted as an "effective mass"   controlling the inertia of the oscillator. Furthermore, one can define a space-dependent effective friction coefficient as   
\begin{equation}
\gamma_{\rm eff}(x)\equiv\gamma \left[1+ (2\pi v_k\tau/L)\cos(2\pi x/L)\right].
\label{eq:sdf}
\end{equation}
Note that for functions which depend only on the spatial variable $x_t$ we often remove the subscript. 
 For large correlation times $\tau \geq L/2\pi v_k$ the effective friction may even take negative values, a feature that can lead to active spontaneous oscillations even for $f=0$~\cite{martin1999active}. When $L\gg 2\pi$ we can approximate $\gamma_{\rm eff}(x)\simeq \gamma \left( 1+ \frac{2\pi v_k\tau }{L}\right) - \frac{4\pi^3 v_k\tau x^2}{L^3} $ and $\sin(2\pi x/L)\simeq 2\pi x/L - (2\pi x/L)^3/6 $, such that Eq.~(\ref{eum}) becomes analogous to a   driven Van~der~Pol oscillator~\cite{belousov2020volterra,belousov2019volterra}.

In what follows we will focus our efforts in  deriving  finite-time statistics of our active-matter model for which analytical calculations are particularly challenging. The active  noise is  characterized by its  strength and also its exponential autocorrelation through its two independent parameters  $\Da$ and $\tau$, which is  a result of the coarse-grained dynamics of the  oscillator within the bacterial bath. The noise parameters $\Da$ and $\tau$ may be experimentally controlled e.g. in a bacterial bath by changing the density of the bacteria or the resources available to them (e.g. food, oxygen, etc.). Interestingly,  increasing the correlation time $\tau$ has two competing effects in the dynamics of the oscillator: a "cooling" effect corresponding to a decrease of strength of the active noise and an enhancement of the persistence in the oscillator's motion.
We will often make use of two limits that  are of particular physical relevance
\begin{itemize}
\item Thermal bath limit ($\tau\to 0$ and $T_{\rm eff}\to T$).  Using the effective Einstein relation~(\ref{eq:Teff}) and the property $\lim_{\epsilon\to 0} \exp(-|x|/\epsilon)/\epsilon=2\delta(x)$ one gets $ \lim_{\tau\to 0}(\gamma^2 \Da/\tau)\exp({-|t-t'|/\tau})=2\gamma k_B T \delta(|t-t'|)$,   and Eq.~(\ref{eum}) becomes a driven overdamped Langevin equation of a Brownian particle in a thermal bath at temperature $T$. Within this limit we assume Einstein's relation $D=k_{\rm B}T/\gamma$. 
\item Deterministic limit ($\Da\to 0$), which leads to a well-known dynamical system~\cite{strogatz2001nonlinear} with a stable fixed point at $  x^*= (2\pi/L)\arccos({\sqrt{1-(f/k)^2}})$ when $0<f/k<1$, whereas when $f/k>1$ the system just keeps on oscillating at a frequency that is dependent on the phase of the oscillator. Such  system undergoes a saddle-node bifurcation at the critical point $f=k$ 
whose vicinity   is characterized by a scaling behaviour with respect to the dynamical quantities. Note  that  $\tau\to\infty$  also corresponds to the deterministic limit, as from Eq.~(\ref{eq:2}) $\eta$ becomes fully correlated while $\Da/\tau\to0$.
\end{itemize}

\subsection{Steady-state power}

 In the presence of the bath, the fluctuations help the oscillator to escape the potential minimum near the fixed point and to complete full rotations (see Fig.~\ref{Fig.1}b, bottom). Such dynamics is characterized by a work input by the external torque on the system, which can be quantified using the methods of stochastic thermodynamics.  Following Sekimoto~\cite{sekimoto2010stochastic},  the fluctuating heat absorbed and work exerted on the oscillator along a trajectory $\{x_s\}_{s=0...t}$ are respectively given by
\begin{eqnarray}
    Q_t&=&\int_{x_o}^{x_t}(-\gamma \dot{x}_{s} + \eta_{s})\circ dx_{s}=(U_t-U_0) - W_t\label{eq:heatSeki}\\
    W_t&=& \int_{0}^{t}f \dot{x_{s}}ds= f(x_t-x_0),\label{eq:workSeki}
\end{eqnarray}
where $U_t = -(k L/2\pi) \cos (2\pi x_t/L)$ is the fluctuating energy of the system at time $t$ and $\circ$ denotes the Stratonovich product.  The definitions~(\ref{eq:heatSeki}-\ref{eq:workSeki}) of stochastic heat and work ensure the fulfilment of the first law of thermodynamics for each individual stochastic trajectory~\cite{sekimoto1998langevin}. 

First we discuss long-time properties of the work exerted on the oscillator. To this end, we realize that its average value is extensive with time $\langle W_t\rangle \propto t$, hence it is interesting to study the power inputted into the system in the stationary limit. 
At the steady state, the average power done by the external torque  is given by
\begin{eqnarray}
\dW &=& f\langle\dot{x}\rangle_{\rm st}\\
&=&  fv_f - f v_k\langle\sin(2\pi   x/ L) \rangle_{\rm st}, \label{eq:11}
\end{eqnarray}
where in the second line we have used Eq.~(\ref{eq:1}) and introduced the steady-state average $\langle F(x)\rangle_{\rm st}\equiv \int \textrm{d}x P_{\rm st}(x)F(x)$ for any real function $F(x)$. One may note that the power is proportional to the particle current, $\dot{x}$, an important theme  in the study of Brownian motors~\cite{reimann2002brownian}.
Since the Langevin equation is non-linear, it is usually easier to work at the level of ensembles  to compute the statistics of  different thermodynamic quantities. Indeed,  the associated Fokker-Planck equation is a linear partial differential equation with non-linear coefficients. 

\subsection{Power exerted in a thermal  bath}
First, we discuss as an appetizer the thermal bath limit $\tau\to0$ and $T_{\rm eff}\to T$ in which  the stationary distribution $P_{\rm st}(x)$ obeys
\begin{equation}
 \frac{\textrm{d}}{\textrm{d}   x}\left[\left(-v_f+v_k\sin{\left(\frac{2\pi   x}{L}\right)}\right)P_{\rm st}(  x)\right]+D\frac{\textrm{d}^2 P_{\rm st}(  x)}{\textrm{d}   x^2}=0.
\end{equation}
which follows from taking $\partial_t P(  x,t)=0$ in the associated Fokker-Planck equation. 
Following Ref.~\cite{stratonovich1967topics} we derive in~\ref{app:a} an exact analytical expression for the average power
\begin{equation}
    \langle \dot{W} \rangle = f\langle\dot{  x}\rangle= \frac{2Df}{L}\sinh\left(\frac{v_f L}{2D}\right)\left|I_{{\rm i}(v_f L/2\pi D)}\left(\frac{v_k L}{2\pi D}\right)\right|^{-2},
    \label{eq:10}
\end{equation}
 Here,  $I_{{\rm i}z}(y)$ is the modified Bessel function of first kind with purely imaginary order, with  $z$ and  $y$ two real numbers.   From Eq.~(\ref{eq:10}) it can be shown that the average power has a power-law behaviour at the critical point ($f\simeq k$) in the limit of $D$ small (see~\ref{app:a}), given by
\begin{equation}
    \langle \dot{W}\rangle=  \frac{3^{4/3} \Gamma(2/3)^2}{(4\pi)^{2/3}}f v_f^{2/3}\left(\frac{D}{L}\right)^{1/3}\simeq 1.47 \left(\frac{f^5 D}{\gamma^2 L}\right)^{1/3}.
    \label{scaling}
\end{equation}

\begin{figure}[ht]
	\centering
	\includegraphics[width=\textwidth]{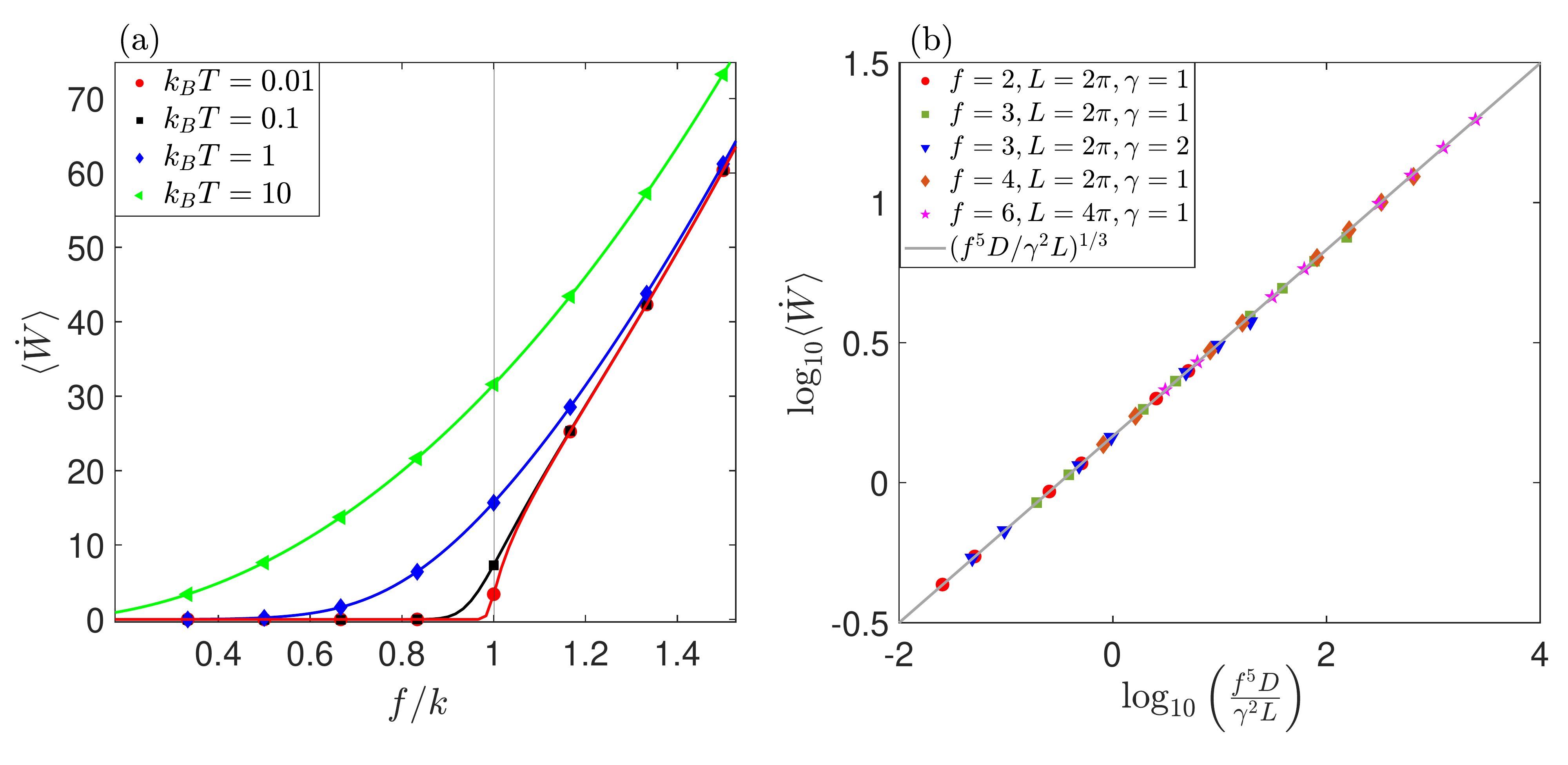}
	\caption{(a) Average power inputted into the oscillator $\dW$ in the thermal bath limit ($\tau=0$ in Eqs.~(\ref{eq:1}-\ref{eq:2})    as function of the scaled driving torque $f/k$ for  different bath temperatures (see legend): numerical  simulations (symbols) and exact analytical expression~(\ref{eq:10}) (lines).  (b) Scaling of the average power  with $(f^5D/\gamma^2L)$ at the critical point $f=k$: numerical simulations (symbols) and Eq.~(\ref{scaling}). In (a) the parameters of the simulations were set to $L=2\pi,\gamma=1,k=6$; in (b) $k=6$. For both panels we performed $10^4$ simulation runs each with  time step $\Delta t=10^{-2}$ and duration $t_{\rm sim}=800$.  }
	\label{average}
\end{figure}

Figure~\ref{average} shows that Eqs.~(\ref{eq:10}-\ref{scaling}) are in excellent agreement with results obtained from numerical simulations of  Eq.~(\ref{eq:1}) in the thermal bath limit, for  a  wide range of parameters that we explore both below and above the critical point.
 When the friction and noise balance, the system is transported at a certain average velocity which   grows with the bath temperature. Even though the fluctuations are symmetric, the asymmetry caused by the external torque   entails an increase of the power with the temperature of the bath  as the noise helps to overcome barriers, until it saturates to $\langle \dot{W} \rangle=f^2/\gamma$ for $T\to\infty$.
The power is a smooth function of $f/k$ everywhere and exhibits a non-analytic behaviour (i.e. discontinuity of its first derivative) in the critical point only in the deterministic limit ($D\to 0$),
 \begin{equation}
    \langle \dot{W}\rangle =
    \left\{  \begin{array}{r@{\quad}cr} 
       0,  & f\leq k,\\
       f\sqrt{v_f^2-v_k^2}, & f>k
    \end{array}\right. \label{eq:15}
\end{equation}
 similarly  to  a second-order phase transition. Note that in this limit, the average  power is  independent of the noise strength but  also of the period length of the potential $L$.
Figure~\ref{average}b confirms with numerical simulations the scaling of the power with the noise strength as $D^{1/3}$.  Note that the scaling  exponent in the thermal bath is dependent on the normal form of the potential~\cite{reimann2001giant}. 




\subsection{Power exerted in an active bath}

\begin{figure}[h!]
	\centering
	\includegraphics[width=0.95\textwidth]{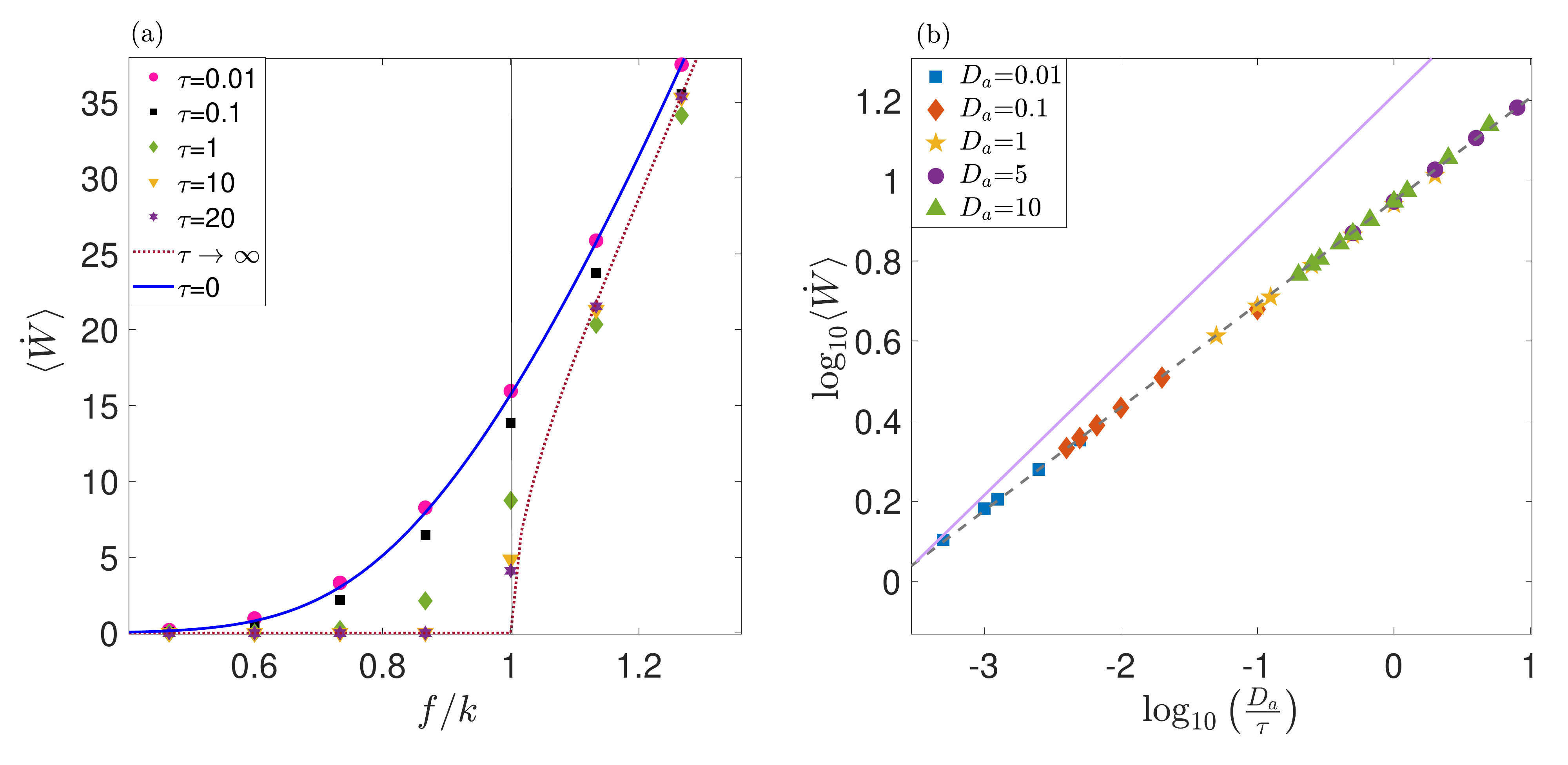}
	\caption{(a) Average    steady-state power exerted on the oscillator Eqs.~(\ref{eq:1}-\ref{eq:2}) as a function of the scaled external torque for different  correlation times of the bacterial bath: simulations (symbols), analytical expression in the thermal bath limit  (Eq.~(\ref{eq:10}), blue solid line) and analytical expression in the deterministic limit (Eq.~(\ref{eq:15}),  red dotted line).  
		Simulation parameters: $L=2\pi,\gamma=1,k=6,\Da=1,\Delta t=10^{-2}$. For $\tau=0.01$ we used $\Delta t=10^{-3}$.   (b) Scaling of the  steady-state power with $\Da/\tau$   for different correlation times $\tau$ and noise strength  $\Da$ of the bacterial bath. The solid purple line is to show the different power law nature of the average power $\langle\dot{W}\rangle\sim D^{1/3}$ in the thermal and  $\langle\dot{W}\rangle\sim D_a^{1/4}$ of the active bath for any fixed $\tau$. Simulation parameters: $\gamma=1,L=2\pi,f=k=2\pi, \Delta t=10^{-2}$. All these simulations are done for a duration of $t=400$ in the steady state. }
	\label{fig:13}
\end{figure}

We now ask the question of how  controlling the correlation time $\tau$ of the active bath impacts on the average power exerted in the oscillator near its critical point. To answer this question we plot in Fig.~\ref{fig:13} results from simulations for the stationary power as a function of the reduced torque $f/k$, for different values of the bath correlation time. Figure~\ref{fig:13}a shows that  the power is bounded from below and above  by its value in the deterministic ($\tau\to\infty$) and thermal bath  ($\tau\to 0$) limits, respectively, for the values of $f$ that we explored. The dependency of the power with $\tau$ for a given value of $f$ is highly non-trivial, and a key result of our work that we explain below in  Sec.~\ref{sec:mpab}.


The presence of non-linear potential gives rise to scaling behavior at the critical point, whose characteristics depend on the form of potential and the properties of the bath.  Our numerical results reveal that the average power has a scaling behaviour as function of strength of the noise, $\langle \eta_t^2 \rangle = \Da/\tau$, for small values of $(\Da/\tau)$ as shown in Fig.\ref{fig:13}b. These results  suggest the scaling behaviour 
\begin{equation}
\langle \dot{W} \rangle \sim \left( \frac{D_{\rm a}}{\tau}\right)^\beta,
\label{eq:16}
\end{equation}
such that the critical exponent, $\beta=0.25\pm 0.01$, which we conjecture to be equal to $1/4$.  We remark that the  proportionality constant accompanying the term $(\Da/\tau)^{1/4}$ may depend on other parameters of the system.  Therefore, we find that the power exerted on the oscillator at its critical point in a bacterial bath displays a slower growth with~$\Da^{1/4}$   than with the diffusion coefficient $D^{1/3}$ when  immersed in a thermal bath. 


\subsection{Minimum power in active baths}
\label{sec:mpab}
In this subsection we provide further details about the dependency of the stationary power with the bacterial bath correlation time both below and above the critical point. We first discuss here results from numerical simulations and compare them with theoretical predictions in the thermal bath and deterministic limit.

\begin{figure}[H]
   \centering
    \includegraphics[width=0.95\textwidth]{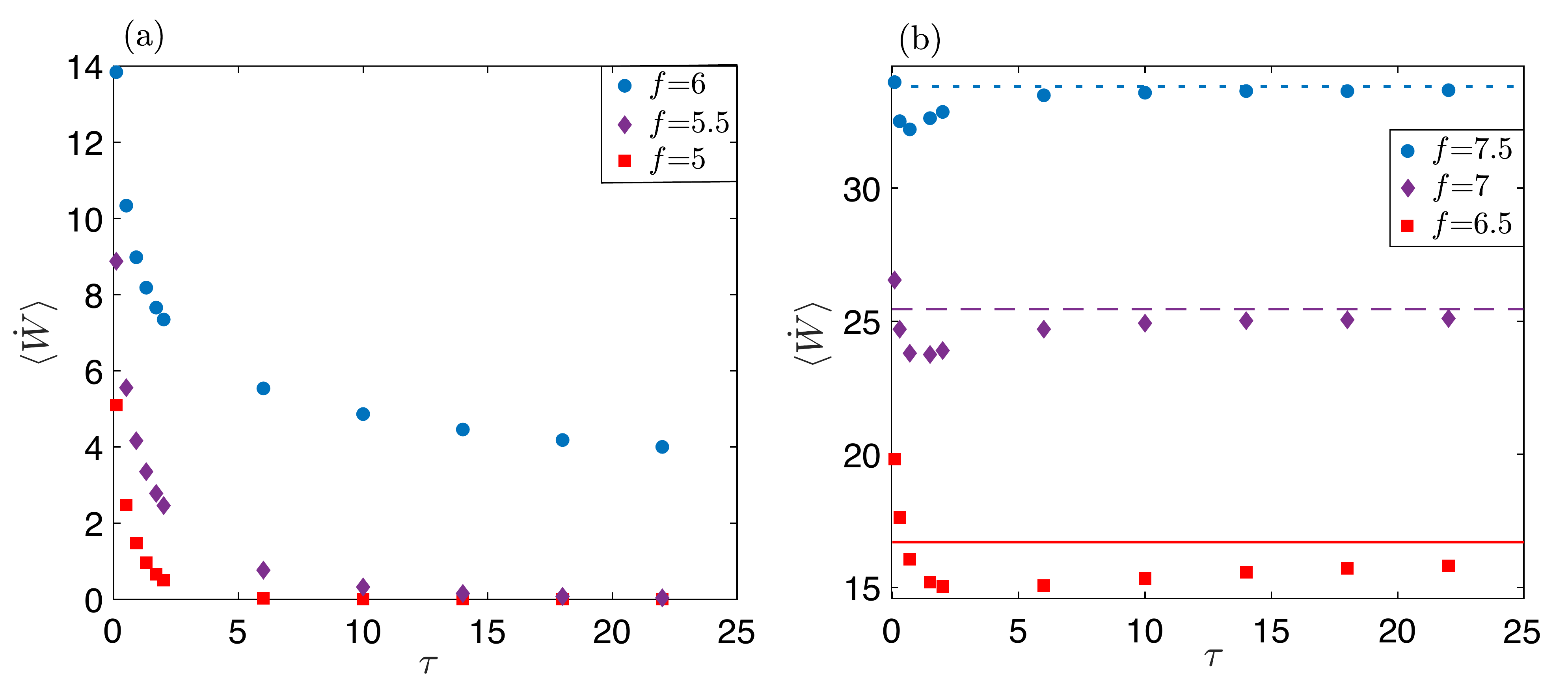}
    \caption{ Results from numerical simulations (symbols) for the average steady-state power exerted on the oscillator given by Eqs.~(\ref{eq:1}-\ref{eq:2}) as a function of the  correlation time of the bacterial bath, for different value of the external torque (see legend):   below the critical point $f<k$ (a),  and above the critical point $f>k$ (b). 
    In (b), the horizontal lines are given by Eq.~(\ref{eq:15}) i.e. the average power in the deterministic limit for different values of $f$ ($f=6.5$ red solid line, $f=7$ purple dashed line, and $f=7.5$ blue dotted line). Parameters of the simulations: $\gamma=1,D_a=1,L=2\pi,k=6$ simulation time step $\Delta t =10^{-2}$  and the averages are computed for $10^{5}$ trajectories each of duration~$t=400$. \label{fig:4}}
\end{figure}

Below the critical point ($f<k$) the power decreases monotonously with the bacterial bath correlation time (see Fig.~\ref{fig:4}a)  from its thermal bath limit to its deterministic limit.  Qualitatively, this behaviour can be captured by a driven oscillator in a thermal bath with a temperature that decreases monotonously with $\tau$.
Hence below the critical point, the increasing  correlation time of the bath  corresponds to an "effective cooling" of the oscillator.

 
Above the critical point $f>k$  the   power displays  a non-monotonous behaviour with the bath correlation time $\tau$ (see Fig.~\ref{fig:4}b). For small correlation times, the average power decreases with $\tau$ to a minimum value that can be  below the deterministic (i.e. zero temperature) limit. We denote $\tau_c$ the optimal bath correlation time at which the power is minimized.  For larger values of $\tau$ the power increases from its minimal value to its deterministic limit. 
The non-monotonous  behaviour of $\langle\dot{W}\rangle$ vs $\tau$ results from a competing effect between the   decreasing strength of the noise at small $\tau$ and the effect of increasing persistence for large $\tau$.
For small persistence times $\tau<\tau_c$, the bacterial bath  has an effective "cooling effect"  on the oscillator thus decreasing the average power exerted on its motion.  For $\tau>\tau_c$ the persistent motion of the bacteria enhance  the current of the oscillator until its deterministic limit for $\tau\to\infty$.



\subsection{Approximation methods}

 A fully analytical solution of the stationary probability density and thus of the power is particularly challenging and not yet available to our knowledge. 
Over the last decades, a series of  approximations have been developed to tackle fluctuations of stochastic systems in the presence of colored noise, see e.g.~\cite{hanggi1995colored,sancho1982analytical,PhysRevA.33.467,PhysRevA.35.3086,PhysRevA.35.4464,PhysRevE.52.159}. 
The presence of an exponential autocorrelation function implies that the Kramers-Moyal expansion does not converge to a standard Fokker-Planck equation, but to the following partial differential equation with a memory kernel~\cite{sancho1982analytical}
 \begin{eqnarray}
\hspace{-1cm}     \frac{\partial P(x,t)}{\partial t}=&-&\frac{\partial}{\partial x}\left[\left(v_f-v_k\sin{\left(\frac{2\pi   x}{L}\right)}\right)P(x,t)\right]\label{eq:fpe}\\&+&\frac{\partial}{\partial x}\left[\frac{\partial}{\partial x} \int_0^t ds \frac{\Da}{\tau} \exp(-|t-s|/\tau) \left\langle \frac{\delta x_t}{\delta \eta_{s}}\Bigg\vert_{x_t=x}\delta(x_t-x)\right\rangle  \right].\nonumber
 \end{eqnarray}
Here, $\delta$ denotes here functional derivative. Note that in the limit $\tau\to 0$ Eq.~(\ref{eq:fpe}) yields the Fokker-Planck equation for a drift-diffusion process where $P(x,t)=\langle\delta(x_t-x)\rangle$ with initial condition $x_0$ at time $t=0$. 
 A useful shortcut to treat Eq.~(\ref{eq:fpe}) is the usage of  Markovian approximations, such that we can write an effective Fokker-Planck equation for the system, with suitable state-dependent drift and diffusion coefficient.  In what follows we will first discuss the application to our model of two popular approximations (Fox's~\cite{PhysRevA.33.467} and unified colored noise approximation UCNA~\cite{PhysRevA.35.4464}) that have been used to describe the equilibrium dynamics of systems in the presence of colored noise, and develop a new approximation scheme valid for large values of $\tau$. 
 

\subsubsection{Small $\tau$ approximations}
\label{sec:FoxUCNA}

Recently, several  approximations have been introduced in the literature to discuss stochastic systems which relax to an nonequilibrium stationary state in the presence of colored noise~\cite{fodor2016far,marconi2017heat,bonilla2019active}. Notably such approximations are suited for systems  where the coarse-grained probability  current (e.g. along the variable $x$) vanishes, i.e. these approximations predict the same stationary distribution $P_{\rm st}(x)$.  A celebrated approximation to Eq.~(\ref{eq:fpe}) was developed by Fox in Ref.~\cite{PhysRevA.33.467} using functional calculus. When applied to our model Eqs.~(\ref{eom1}-\ref{eom}) Fox's approximation yields   
\begin{equation}
\fl     \frac{\partial P_{\rm Fox}(  x,t)}{\partial t}=-\frac{\partial}{\partial   x}\Bigg\{[v_f-v_k\sin(2\pi  x/L)]P_{\rm Fox}(  x,t)\Bigg\} + \Da \frac{\partial^2}{\partial   x^2} \frac{P_{\rm Fox}(  x,t)}{\left[1+(2\pi v_k\tau/L)\cos\left(\frac{2\pi  x}{L}\right)\right]},
     \label{eqn:Fox}
\end{equation}
where $P_{\rm Fox}(x)$ stands for the propagator of an auxiliary Markovian process described by the Fokker Planck equation~(\ref{eqn:Fox}). Such dynamics  follows a heterogeneous diffusion process whose statistics converge to that of the actual process $P_{\rm Fox}(x,t)\simeq P(x,t)$ in the limit of $\tau$ small.  
On the other hand, the so-called unified colored noise appoximation (UCNA) provides an accurate description both in the limit of $\tau$ small and $\tau$ large, for  Langevin systems moving in conservative potentials.  UCNA is formulated in the $(x,\eta)$ phase space where the dynamics is underdamped. Rescaling time $\tilde{t}=t\tau^{-1/2}$, and making an adiabatic approximation (i.e. neglecting inertial contributions)  for $\tau\to 0$ and $\tau \to \infty$ yields an overdamped Langevin process whose corresponding Fokker-Planck equation  is given by
\begin{eqnarray}
\fl
     \frac{\partial P_{\rm UCNA}(  x,t)}{\partial t}&=&-\frac{\partial}{\partial   x}\left[\frac{v_f-v_k\sin{\left(2\pi  x/L\right)}}{1+(2\pi v_k\tau/L)\cos\left(2\pi  x/L\right)}\,P_{\rm UCNA}(  x,t)\right] \nonumber\\ &+&\Da\frac{\partial}{\partial   x}\frac{1}{\left[1+(2\pi v_k\tau/L)\cos\left(2\pi  x/L\right)\right]}\frac{\partial }{\partial   x} \frac{P_{\rm UCNA}(  x,t)}{\left[1+(2\pi v_k\tau/L)\cos\left(2\pi  x/L\right)\right]}.
     \label{Eqn:ucna}
\end{eqnarray}
A  key difference between Fox's and UCNA approximations is in the probability current $J(x,t)$, where UCNA has overall rescaling with respect to Fox as  $J_{\rm UCNA}(x,t)=\left[1+\frac{2\pi v_ k\tau}{L}\cos\left(\frac{2\pi  x}{L}\right)\right]^{-1} J_{\rm Fox}(x,t)$.  This implies that even though the two approximations lead to the same   equilibrium distribution, their corresponding nonequilibrium stationary states have distinct   distributions. We remark that the regime of applicability of Fox's approximation and UCNA is restricted to values of $\tau$ such that the effective  space-dependent  diffusivity in Fox's and the space-dependent viscosity in UCNA are positive, i.e. when  $\tau<\tau^{\star}\equiv L/2\pi v_k$.  
 
 We now apply UCNA and Fox's approximation to obtain estimates of the steady-state power in our active-matter model. To this aim, we evaluate the steady-state average in the definition of the power, see Eq.~(\ref{eq:11}), by averaging $\sin(2\pi x/L)$ with the numerical  stationary solution of Eqs.~(\ref{eqn:Fox}) and~(\ref{Eqn:ucna}), yielding the estimates $\langle\dot{W}\rangle_{\rm Fox}$ and $\langle\dot{W}\rangle_{\rm UCNA}$, respectively. We investigate the dependency of these estimates with $\tau$ above the critical point and compare them with the simulation results reported in the previous section. 
 \begin{figure}[t]
\centering
\includegraphics[width=\textwidth]{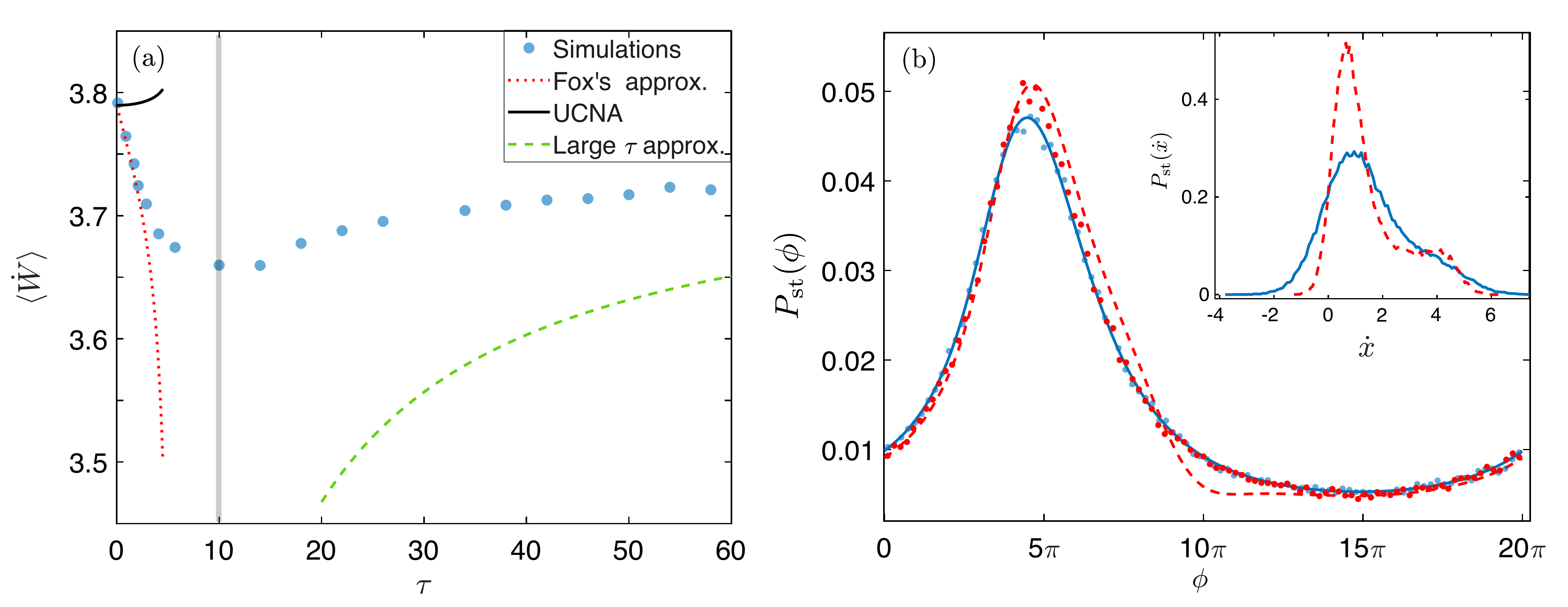}
\caption{(a) Average steady-state power inputted into the system given by Eqs.~(\ref{eq:1}-\ref{eq:2}) as a function of the correlation time of the bacterial bath: simulations (blue circles), theoretical  approximations UCNA $\langle\dot{W}\rangle_{\rm UCNA}$ (black solid line), and Fox's $\langle\dot{W}\rangle_{\rm Fox}$  (red dotted line), and large-$\tau$ approximation $\langle\dot{W}\rangle_{\rm LT}$ given by Eq.~(\ref{eqn:largetau}) (green dashed line). The vertical line  is set to $\tau^{\star}= L/(2\pi v_k)$ where both  approximations become nonphysical. (b)  Stationary distribution of the phase $\phi=x\,\textrm{mod}\, L$ for different  correlation times of the bath: $\tau=1$ (simulations blue circles, Fox's approximation blue solid line) and $\tau=4.5$ (simulations red circles, Fox's approximation red dashed line). Inset: probability density of the finite-time velocity obtained from simulations for $\tau=1$ (blue solid line) and for $\tau=4.5$ (red dashed line).  
Parameters of the simulations: $f=2.5,k=2,\gamma=1,\Da=1,L=20\pi$. The instantaneous velocity was computed using first order finite difference method in the simulations.}
\label{fig:9}\end{figure}
Figure~\ref{fig:9}a shows  $\langle\dot{W}\rangle_{\rm UCNA}$ increases monotonously for  $\tau$  and hence fails to predict the behaviour of average power (further numerical evidence can be found in \ref{app:c}). On the other hand,  $\langle\dot{W}\rangle_{\rm Fox}$  reproduces the decreasing behaviour of the power, and an excellent estimate, for small values of the correlation times at which the stationary distribution of the velocity is close to Gaussian (see Fig.~\ref{fig:9}b).  Both UCNA and Fox's approximation are unable to predict the power optimization at finite correlation times, which occurs at a time beyond their range of applicability for the example in Fig.~\ref{fig:9}. Notably we find that the non-monotonous behaviour in the average power is associated with the non-Gaussian distribution of the angular velocity $\dot{x}$ which cannot be captured by Fox's approximation and UCNA unless one considers  higher-order contributions in both $(  x,\dot{  x})$ to capture it~\cite{marconi2017heat,bonilla2019active}.
 
 
\subsubsection{Large $\tau$ approximation:}

The Fokker-Planck equation associated with the Markovian two-dimensional  Langevin Eqs.~(\ref{eom1}-\ref{eom}) in the $( x,\eta)$ space is given by
	\begin{equation}
	\frac{\partial P }{\partial t}=-\frac{\partial }{\partial   x}\Big[\left(v_f-v_k\sin(2\pi  x/L) +\eta\right)P\Big] +\frac{\partial}{\partial \xi}\left[\left(\frac{\eta}{\tau}+\frac{\Da}{\tau^2}\frac{\partial}{\partial \eta}\right)P\right],\label{eq:21}
	\end{equation}
	where $P\equiv P(x,\xi,t)$ is the probability density to be at $x_t=x$, and $\xi_t=\xi$ at time $t$.
In the large correlation time limit, we may neglect the second term in the right hand side of Eq.~(\ref{eq:21})   to solve  the stationary large-$\tau$ (LT) marginal distribution
\begin{equation}
 \fl   P_{\rm LT}(  x)= \mathcal{N}^{-1}\left[\frac{1}{v_f-v_k\sin{\left(2\pi  x/L\right)}}+ \frac{1}{2}\left(\frac{2\Da}{\tau}\right)\frac{1}{\left[v_f-v_k\sin{\left(2\pi  x/L\right)}\right]^3}+O[(\Da/\tau)^3]\right],
    \label{eq:21}
\end{equation}
where $\mathcal{N}$ is a normalization factor that depends on the order to which the sum in Eq.~(\ref{eq:21}) is truncated. 
Taking terms up to order $2\Da/\tau$ in $P_{\rm LT}(x)$, we derive in~\ref{app:b} a closed form for  average power inputted into the system in the large-$\tau$ approximation, given  by,
\begin{equation}
    \langle \dot{W}\rangle_{\rm LT} =\frac{\tau f (v_f^2-v_k^2)^{5/2}}{(v_f^2+v_k^2/2)\Da+\tau(v_f^2-v_k^2)^2}.   
\label{eqn:largetau}
\end{equation}
Note that Eq.~(\ref{eqn:largetau})  converges  at large $\tau$  to the value of the power in the deterministic limit above the critical point of the oscillator, cf. Eq.~(\ref{eq:15}). 
 
Figure~\ref{fig:9}a shows that the  approximation given by Eq.~(\ref{eqn:largetau}) converges from below, at large $\tau$, to the value of the power obtained from numerical simulations. Even though for the parameters analyzed $\langle \dot{W}\rangle_{\rm LT} $ provides a loose lower bound for intermediate values of $\tau$, it captures the increasing behaviour of the average power at large  $\tau$ which cannot be reproduced using UCNA and Fox's approximation. 

\section{Fluctuations and Uncertainty relations}
\label{sec:3}
At the  mesoscopic scale, fluctuations play a key role in determining the phenomenology of driven oscillators~\cite{wang2000coherence,matsumoto1983noise,gammaitoni1998stochastic}. Recent works have established a thermodynamic relation between the minimum rate of dissipation needed to achieve a desired amount of uncertainty within the so-called {\em thermodynamic uncertainty relations}~\cite{barato2015thermodynamic,horowitz2019thermodynamic,hasegawa2019fluctuation,pigolotti2017generic,di2018kinetic,dechant2018current}. 
Motivated by these recent works, we evaluate in this section the fluctuations (variance and Fano factor) of stochastic work exerted to the oscillator in our active-matter model.  




To explore the uncertainty in the fluctuations of the work exerted on the oscillator, we evaluate the variance of the stochastic work  $ {\rm Var}(W_{t})=f^2(\langle x^{2}_t\rangle_{\rm st}-\langle x_t\rangle_{\rm st}^2)$. 
In the thermal bath limit,  variance of the work inputted into the system can be computed in the limit  large observation times  $t$ following Ref.~\cite{reimann2001giant}, which yields
\begin{equation}
    \textrm{Var}(W_t) \simeq  2f^2Dt \frac{\int_0^{L} \frac{d  x}{L} I_+^2(  x)I_-(  x)}{\left[ \int_0^{L} \frac{d  x}{L} I_+(  x)\right]^3},
    \label{eq:23}
\end{equation}
where 
\begin{equation}
\fl I_{\pm}(x)=\int_0^L\frac{dx'}{D}\exp{\left[\left(\mp kL\cos{\left(\frac{2\pi  x}{L}\right)} \pm kL\cos{\left(\frac{2\pi  (x \mp x')}{L}\right)}- 2\pi f x\right)/2\pi k_BT\right]}.\label{eq:defIs}
\end{equation}
Equation~(\ref{eq:23}) reveals the phenomenon of   "giant acceleration" of free diffusion~\cite{reimann2001giant} i.e. an enhanced effective diffusion coefficient above the bare diffusivity $D$ occurring near the critical point of the oscillator. For the case of an oscillator within an active bacterial bath, an analytical expression for the work variance is yet not available to our knowledge.

\begin{figure}[h!]
    \centering
    \includegraphics[width=0.95\textwidth]{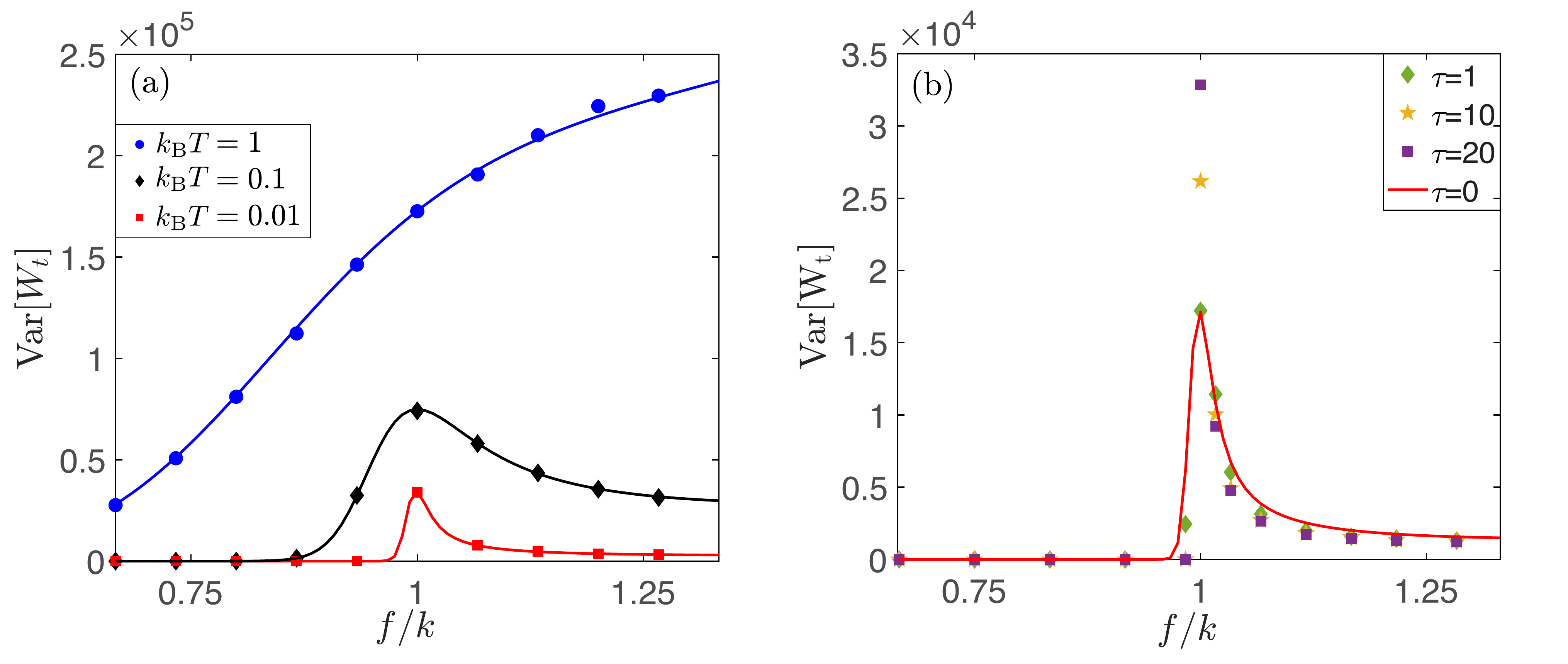}
    \caption{(a)  Variance in work exerted into the oscillator immersed in a thermal bath as function of scaled driving torque for different temperatures (see legend): simulations (symbols) and analytical expression   Eq.~(\ref{eq:23}). 
    Parameters: $\gamma=1,L=2\pi,k=6$ and observation time $t=800$.  (b) Variance of the work exerted on the oscillator in an active bacterial bath as function of scaled driving torque for different bath correlation times. Parameters: $\gamma=1,L=2\pi,k=6,\Da=0.01, \Delta t=10^{-2}$ and observation time  $t=400$.}
    \label{Fig4}
\end{figure}

We analyze the dependency of the work variance with the external torque near the critical point of the oscillator in a thermal bath  and an active baterial bath in Fig.~\ref{Fig4}a and Fig.~\ref{Fig4}b respectively. In the thermal bath limit, the work variance develops a peak near the critical point for small temperatures. This result is a signature of the underlying bifurcation of the non-linear dynamical system, as  near the bifurcation point, there is an equal contribution of  the variances from oscillating and  trajectories stalled at the fixed point.
 By  increasing the bath temperature  the work variance increases and its peak value is attained at larger values of the  external  torque. 
 We also find a scaling behaviour $\textrm{Var}(W_{t})\sim D^{1/3}$ for fixed observation times (data not shown) as expected from Eq.~(\ref{eq:23}). 
Notably,  in the deterministic (i.e. zero-temperature) limit, the variance is zero for all parameter values. 

Within the bacterial bath, the finite-time work variance presents a dependency with the torque that depends strongly on the correlation time of the bath, see Fig.~\ref{Fig4}b. For the parameter values that we explored, we find that below and above the critical point the work variance is below its thermal bath limit, for any value of the bath correlation time $\tau$. On the contrary, the work variance is  enhanced at the oscillator's critical point when increasing $\tau$ which is accompanied by a sharpening of the  peak  as a function of the scaled torque. This is one of key differences between going to the deterministic case, through the two limits: $\Da\to0$ (Fig.~\ref{fig:4}a) and $\tau\to\infty$~(Fig.~\ref{fig:4}b). 
We rationalize such variance enhancement as a synergy between  the persistence of the motion induced by the bacterial bath and the   "giant acceleration"  of free diffusion near the critical point. 

For a precise motion of a fluctuating oscillator in the form of a "clock"~\cite{barato2016cost} it is desirable that its time-integrated current $x_t$ and hence $W_t$ display low values of uncertainty, which is associated with a minimal thermodynamic cost i.e. entropy production, following the so-called thermodynamic uncertainty relations. We now discuss the tradeoff between cost and precision by analyzing the value of the Fano factor of the stochastic work $F_{W_t}=\textrm{Var}(W_{t})/\langle W_t \rangle$~\cite{manikandan2018exact,barato2015universal,hyeon2017physical}. 
In the thermal bath limit, we have that $\langle W_t\rangle=\langle \dot{W}\rangle t$ and also $\textrm{Var}(W_t)\propto t$ at large times. Hence for large $t$, the Fano factor of the work in the thermal bath limit reaches a saturating, finite value given by
\begin{equation}
    F_{W_t}    =2 f D\frac{\displaystyle\int_0^{L} \frac{d  x}{L} I_+^2(  x)I_-(  x)}{\left(1-\exp[{-fL/k_{\rm B}T}]\right) \left[\int_0^{L} \frac{dx}{L} I_+(  x)\right]^2},
    \label{eq:24}
\end{equation}
where the functions $I_{\pm}(x)$ were defined in Eq.~(\ref{eq:defIs}). 
\begin{figure}[h!]
	\includegraphics[trim=0 0 0 30, clip,width=\textwidth]{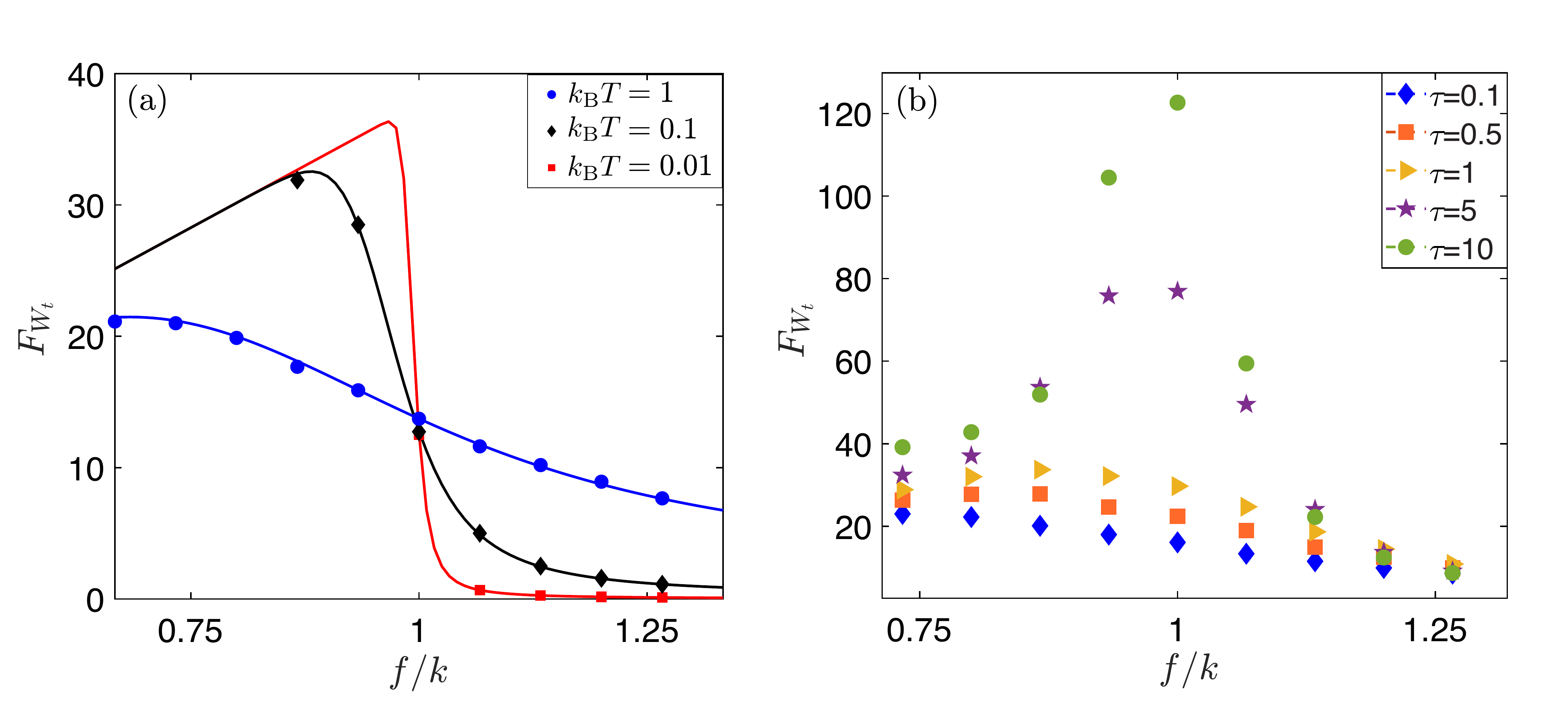} 
	\caption{(a) Finite-time Fano factor $F_{W_t}=\textrm{Var}(W_t)/\langle W_t\rangle$ for the work $W_t$ exerted on the oscillator in a thermal bath as a function of the scaled driving torque for different values of the bath temperature (see legend). (a) Finite-time Fano factor  for the work  exerted on the oscillator in a bacterial  bath as a function of the scaled driving torque for different values of the bath correlation time. In (a) and (b) the symbols are obtained from numerical simulations with parameters: $\gamma=1,L=2\pi,k=6,t=800,\Delta t=10^{-2}$ for (a), and  $\gamma=1,L=2\pi,k=6,\Da=1,t=400,  \Delta t=10^{-2}$ for (b). The solid line in panel (a) is obtained from the analytical expression given by Eq.~(\ref{eq:24}).}
	\label{fig:3}
\end{figure}
Figure~\ref{fig:3}a shows  that the oscillator has highest degree of uncertainty (Fano factor), in a thermal bath, for  values  of the torque below the critical point, and develops a local maximum for values of $f$ converges to the critical point in the deterministic limit. 
Far above the bifurcation point the oscillator  obeys the thermodynamic uncertainty relation, $F_{W_t}\ge 2k_{\rm B}T$~\cite{barato2015thermodynamic,gingrich2016dissipation} for all values of the bath temperature. We also find that the bound is saturated in the drift-diffusion limit of the system, i.e. when $f\gg k$ and the non-linearity of the potential is negligible. 

\begin{figure}[h!]
	\centering
	\includegraphics[width=0.65\textwidth]{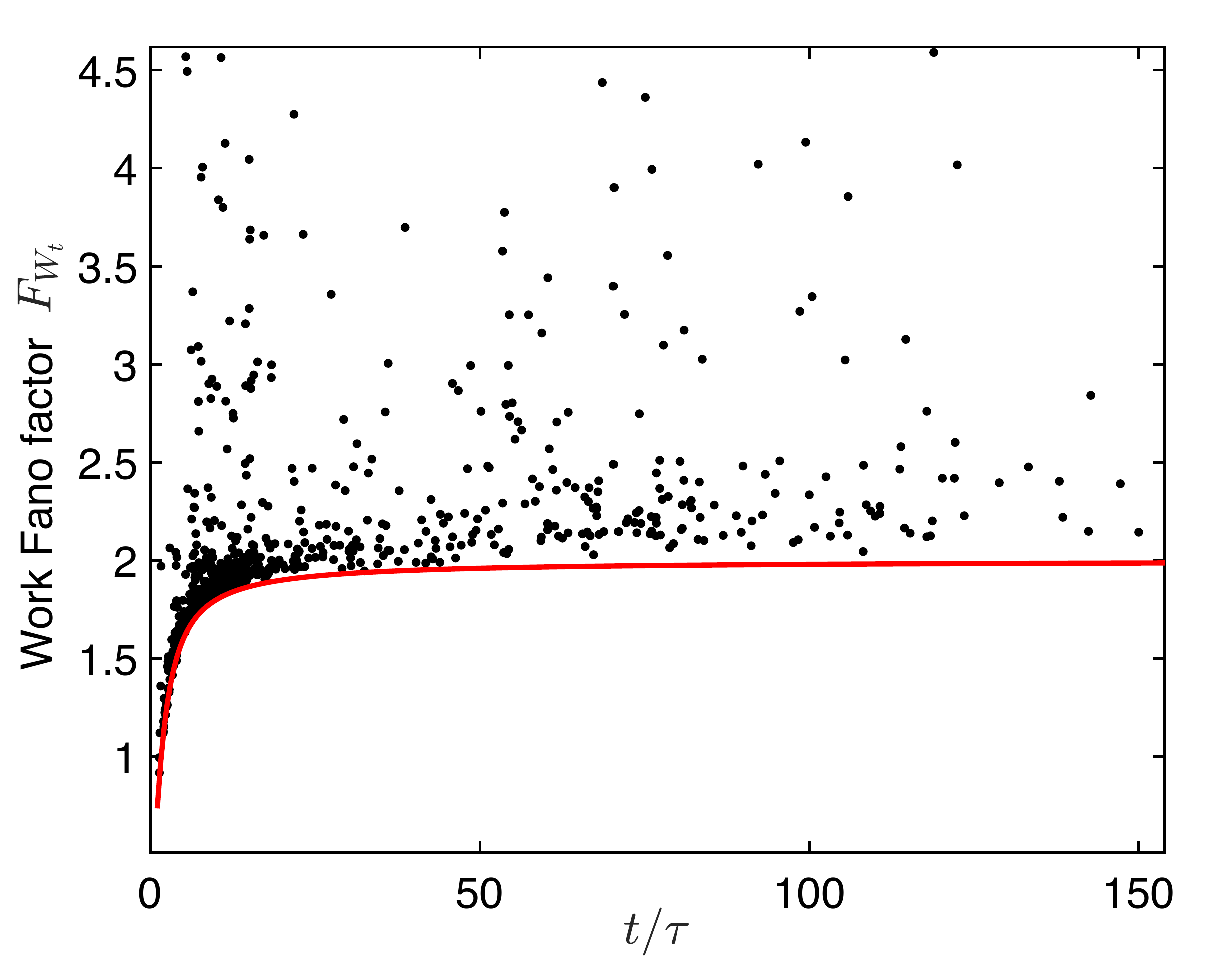} 
	\caption{Numerical verification of the thermodynamic uncertainty relation for active matter~(\ref{Eqn:fanoactive}) above the critical point $f>k$ of the oscillator. 
	 The points correspond to $\sim 600$  values of $f,t$ and $\tau$ randomly picked  from the range $f\in [8,70]$, $t\in [1,40]$ and $\tau\in[50,200]$ respectively. The solid red line is given by the right-hand side of Eq.~(\ref{Eqn:fanoactive}).
		Parameters: $\gamma=1,L=2\pi,k=6,\Da=1, \Delta t=5\times 10^{-3}$. Each of the averages were computed from $2\times 10^{4}$ trajectories.}
	\label{fig:14}
\end{figure}

 In the presence of an active bacterial  bath, the  work variance peaks below the critical torque, see Fig.~\ref{fig:3}b. When increasing the correlation time of the bath, the  optimal value of the torque that leads to maximum work variance   shifts towards the critical point, and the power peak becomes sharper. 
 Notably, we find that the finite-time work  Fano factor obeys the following thermodynamic uncertainty relation for active matter 
\begin{equation}
    F_{W_t}\ge 2k_{\rm B}T_{\rm eff}\left[1-\frac{1-\exp(-t/\tau)}{t/\tau}\right],
    \label{Eqn:fanoactive}
\end{equation}
which we test numerically in  Fig.~\ref{fig:14} for a broad range of system parameters.  Note that the right-hand side of the bound~(\ref{Eqn:fanoactive})  equals to the work Fano factor in the limit $f\gg k$ where the oscillator dynamics is a   drift-diffusion process in the presence of colored noise (see~\ref{app:d}). Figure~\ref{fig:14} reveals that   a precise  input of power in the system  with $F_{W_t}<2k_{\rm B}T_{\rm eff}$ can be achieved within an active bath  at large torque $f>k$ and/or observation times comparable or smaller than the bath correlation time $t<\tau$.  


\section{Connection to experiments: minimal power in bacterial baths}
\label{sec:4}

Finally we draw a  swift  connection to experiments by providing   predictions using  numerical simulations of our model for parameter choices that are close to previously published experimental works. Gomez-Solano {\it et al.} performed experiments with driven colloidal particles  in water trapped with a toroidal optical trap~\cite{gomez2009experimental,gomez2011fluctuations}. In such experiments, the dynamics of the phase was accurately described by Langevin overdamped system subject to  a non-linear  optical force and to thermal noise, similarly to our model given by Eqs.~(\ref{eq:1}-\ref{eq:2}) in the thermal bath limit. On the other hand, Refs.~\cite{maggi2014generalized,argun2016non,wu2000particle} reported experiments studying the diffusion of  micron-sized spherical particles in a film filled with {\it E. Coli} bacterial cells harvested from culture media. In particular, the mean-squared displacement  of polystyrene beads moving  in a freely-suspended bacterial film was accurately described with an instantaneous friction memory kernel and Gaussian colored noise. 

\begin{figure}[h!]
\centering
\includegraphics[trim=0 0 0 0, clip, width=\textwidth]{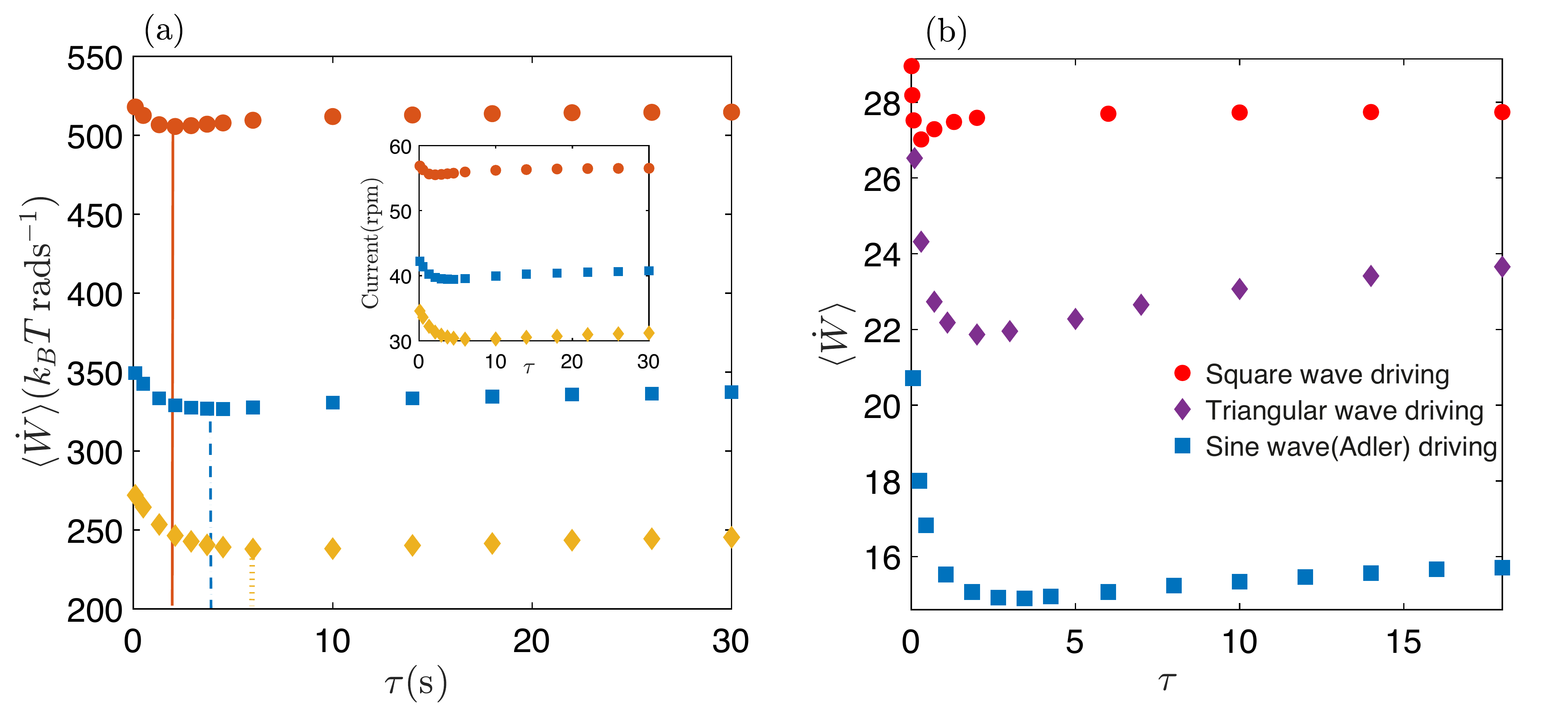} 
\caption{(a) Average stationary power inputted into the oscillator in a bacterial bath, described by Eqs.~(\ref{eom1}-\ref{eom}),  as a function of the bath correlation time. 
The inset shows the  average stationary current mean current (in revolutions per minute, $\rm{rpm}$) as function of correlation time. The results are obtained from numerical simulation of Eqs.~(\ref{eq:1}-\ref{eq:2}) for torque values $f=0.95 \rm{rad}\, s^{-1}$ (yellow diamonds), $f=1 \rm{rad}\, s^{-1}$ (blue squares),  and  $f=1.1 \rm{rad}\, s^{-1}$ (red circles). The vertical bars are set to the optimal correlation time at which the average power is minimal ($f=0.95 \rm{rad}\, s^{-1}$ solid, $f=1 \rm{rad}\, s^{-1}$ dashed,  and  $f=1.1 \rm{rad}\, s^{-1}$ dotted).  Parameters: $k=0.87 \textrm{rad s}^{-1}, \Da=0.63 \mu \textrm{m s}^{-2}$.  (b) Average power as a function of the bath correlation time  for different driven non-linear periodic potentials, see Sec.~\ref{sec:5} for details. Parameters: $\gamma=1,L=2\pi,k=6, f=6.5,\Da=1,  \Delta t=10^{-2}$. }
\label{fig:7+10}
\end{figure}

 Using knowledge from the aforementioned experimental work~\cite{gomez2011fluctuations,wu2000particle}, we  provide some numerical  predictions about the power done on a driven colloidal particle trapped in a toroidal optical tweezer and embedded in a bacterial bath, which yet has not been realized experimentally to our knowledge.  
We consider a spherical silica  particle of radius $r=1\mu{\rm m}$ immersed in water at $293$K with friction coefficient $\gamma=0.16\, {\rm pNs}/\mu {\rm m}$, which is optically trapped in a toroidal potential of major radius $R=4.12\,\mu {\rm m}$. 
 The strength of the optical potential is $v_k=3.6\, {\rm rad} \mu{\rm m s}^{-1} $.  
  Because we are interested in values of external torque above the critical point we choose $v_f = 4.2\, {\rm rad s}^{-1}$, cf. Eqs.~(\ref{eom1}-\ref{eom}). 
  For the active bacterial bath, we took parameter values from  Ref.~\cite{wu2000particle}, corresponding to a microscopic particle immersed in an {\it E.Coli} bacterial bath. More precisely, we used   $\Da=0.63\, \mu {\rm m s}^{-2}$ and  correlation times of the orders of seconds.  
 For this parameter values, the average power inputted into  the particle is of the order of $\sim 100 k_{\rm B}T/{\rm s}$ above the critical point, and the optimal correlation time at which the power is minimized is of the order of seconds, see Fig.~\ref{fig:7+10}a. Interestingly, we also report estimates of the stationary angular current (Fig.~\ref{fig:7+10}a inset) which can be of the orders of $60$rpm (i.e. one revolution per second) for moderate values of the external torque, which highlights the possibility of constructing {\em bacterial clocks} at the microscale.

\section{Discussion}
\label{sec:5}

We have shown that the presence of correlations in the noise induced by an active (e.g. bacterial) bath  has measurable signatures in the nonequilibrium fluctuations of the work exerted on a non-linear oscillator. In this paper we focused on Langevin systems given by Eqs.~(\ref{eq:1}-\ref{eq:2}) which we used to describe  the motion of  a driven colloidal particle trapped in a periodic potential within a bacterial bath of \textit{E. Coli}. Variants of this model have also been investigated in the context of   self-propelled microswimmers confined in a ring,  hearing by active mechanosensory  hair bundles,  and fluctuations of     Josephson junctions. 

Our numerical results show that the power fluctuations depend strongly on the correlation time of the active bacterial bath $\tau$ below, at, and above the critical point of the oscillator.  Below the critical point, we have found that the average power at steady state is bounded, for all parameter values, from below by its zero-temperature (i.e. deterministic) limit and from above by its value in the limit of thermal fluctuations (i.e. when recovering fluctuation-dissipation).  At the critical point, our simulations lead us to conjecture the power-law behaviour $\langle\dot{W}\rangle \sim (\Da/\tau)^{1/4}$, see Eq.~(\ref{eq:16}), which is genuinely different to the scaling in the thermal bath limit $\langle\dot{W}\rangle \sim D^{1/3}$. Moreover we  found that the scaling of the stationary current in the bacterial bath with the force is  $\langle \dot{x}\rangle_{\rm st} \sim f^{\epsilon(\tau)}$, with $\epsilon(\tau)\leq 2/3$  a $\tau$-dependent exponent for which $\epsilon(0)=2/3$ in the thermal bath limit.

Furthermore, we have unveiled  a very rich phenomenology of such active-matter oscillator above its critical point. In particular, 
both Fox's and Unified Colored Noise (UCNA) approximations fail in describing the dependency of the average power as a function of the bath correlation time. Both approximations are unable to explain the key result of this work: the appearance of a minimum in the curve $\langle \dot{W}\rangle$ vs $\tau$, which takes place at a correlation time that leads to negative effective friction coefficient.  Strikingly, this result establishes that  there exist an optimal correlation time $\tau^{\star}$, near which the average power is  below the power done in the oscillator in the zero-temperature limit.
By developing a new approximation valid at large $\tau$ we have established a new closed form of a lower bound for the power valid at large times and above the oscillator's critical point, see Eq.~(\ref{eqn:largetau}). 
Moreover, we have established a link between the variance and Fano factor of the work and thermodynamic uncertainty relations, and found a finite-time uncertainty relation for active matter~(\ref{Eqn:fanoactive}) that we tested valid in simulations above the critical point. 
Our bound provides further insights on the violation of the Markovian uncertainty relation  as a result of the non Markovian footprints  in the dynamics of active systems~\cite{saryal2019thermodynamic,di2020thermodynamic}.

It remains an open question to investigate  whether the power optimization  with the bath correlation time can be further discussed in relation with the  frenetic aspects of nonequilibria~\cite{maes2020frenesy,basu2015nonequilibrium,baiesi2010nonequilibrium,roldan2019exact}, and whether 
   this result applies to a broader class of systems. To serve an appetizer to this challenging task, we show in  Fig.~\ref{fig:7+10} that the power minimization also occurs for different forms of non-linear periodic potential. 
 Hence, we hypothesize that  the ingredients needed for this phenomenon are the following: (i) a non-linear periodic potential, (ii) external driving and (iii) an active bath. It will be interesting  in the future to study the above phenomena in the context of Brownian motors~\cite{reimann2002brownian,astumian2002brownian} bacterial ratchets~\cite{vizsnyiczai2017light} and many-particle systems with long-range interactions in the onset of synchronization~\cite{gupta2017world}.

\section{Acknowledgements}
This work is part of MUR-PRIN2017 project "Coarse-grained description for non-equilibrium systems and transport phenomena (CO-NEST)"  No. 201798CZL whose partial financial support is acknowledged. A.G. thanks The Abdus Salam International Centre for Theoretical Physics (ICTP) and Scuola Internazionale Superiore di Studi Avanzati (SISSA) for the hospitality and financial support. We acknowledge Sarah Loos and Matteo Marsili for a careful reading of our manuscript. We thank Juan M. R. Parrondo, Andrea Puglisi, Lorenzo Caprini, Christopher Jarzynski, Roman Belousov, Matteo Polettini, Etienne Fodor,  Yuzuru Sato, and Jacopo Grilli for stimulating discussions.

\appendix
%
%
\section{Derivation of average power in the thermal bath limit and its scaling form at the critical point}
\label{app:a}

\underline{Derivation of Eq.~(\ref{eq:10})}. In the thermal bath limit, the oscillator dynamics is described by the  one-dimensional  Langevin equation
\begin{equation}
    \gamma \dot{x}_t=f-k\sin{\left(\frac{2\pi  x}{L}\right)}+\sqrt{2\gamma k_B T} \;\xi_t 
    \label{eq:a1}
\end{equation}
where $\xi_t$ is Gaussian white noise with zero mean and autocorrelation $\langle\xi_t\xi_{t'}\rangle=\delta(t-t')$.
The Fokker-Planck  equation associated with Eq.~(\ref{eq:a1}) is
\begin{equation}
    \frac{\partial \hat{P}( x,t)}{\partial t}=-\frac{\partial}{\partial  x}\left[\left(v_f-v_k\sin{\left(\frac{2\pi  x}{L}\right)}\right)P( x,t)\right]+D\frac{\partial^2 \hat{P}( x,t)}{\partial  x^2}\quad.
    \label{eq:a2}
\end{equation}
where $\hat{P}(x,t)= \sum_{n=-\infty}^{\infty}P(x+nL,t)$ is the reduced probability density, which obeys the normalization condition $\int_0^L \hat{P}( x,t) d x =1$ and $P(x,t)=P(x\pm L,t)$ for all  $x$ and all $t>0$.
In the steady state, the probability current
\begin{equation}
    \hat{J}(x)=\left(v_f-v_k\sin{\left(\frac{2\pi  x}{L}\right)}\right)\hat{P}_{\rm st}( x)-D \frac{\partial \hat{P}_{\rm st}( x)}{\partial  x},
    \label{eq:a3}
\end{equation}
is constant and independent of  $x$.
The  solution of Eqn.~(\ref{eq:a2}) satisfying the aforementioned boundary conditions is \begin{equation}
    \hspace{-1cm} \hat{P}_{\rm st}( x)=C_1\exp\left[a x+b\cos{\left(\frac{2\pi  x}{L}\right)}\right]\int_{-\infty}^ x \exp\left[-\left(a\phi+b\cos{\left(\frac{2\pi \phi}{L}\right)}\right)\right]d\phi
 \end{equation}
where the constants
\begin{equation}
    a\equiv\frac{v_f}{D}, \quad  b\equiv\frac{v_k L}{2\pi D}.
\end{equation}
Now rearranging the above formula ,
\begin{eqnarray}
&&\hspace{-2cm}\hat{P}_{\rm st}( x)=
    C_1 \exp\left[L a\right] \exp\left[a x+b\cos\left(\frac{2\pi  x}{L}\right)\right]\int_{-\infty}^{ x+L} \exp\left[-\left(a\phi+b\cos\left(\frac{2\pi \phi}{L}\right)\right)\right]d\phi \nonumber\\
   &&\hspace{-2cm}=  C_1 \exp\left[L a\right] \exp\left[a x+b\cos\left(\frac{2\pi  x}{L}\right)\right]\left(\sum_{n=0}^\infty\int_{x-nL}^{ x-(n-1)L} \exp\left[-\left(a\phi+b\cos\left(\frac{2\pi \phi}{L}\right)\right)\right]d\phi\right) \nonumber\\
     &&\hspace{-2cm}= C_1 \exp[La]\left(\sum_{n=0}^{\infty} \exp(nLa)\right) \exp\left[a x+b\cos\left(\frac{2\pi  x}{L}\right)\right]\int_{ x}^{ x+L} \exp\left[-\left(a\phi+b\cos\left(\frac{2\pi \phi}{L}\right)\right)\right]d\phi\nonumber\\
     &&\hspace{-2cm}=\frac{1}{N} \exp\left[a x+b\cos\left(\frac{2\pi  x}{L}\right)\right]\int_{ x}^{ x+L} \exp\left[-\left(a\phi+b\cos\left(\frac{2\pi \phi}{L}\right)\right)\right]d\phi, \label{eq:a6} \hspace{-2cm}
\end{eqnarray}
where $N$ is defined as
\begin{equation}
  N=-\frac{(\exp\left[-L a\right]-1)}{C_1}.
\end{equation}
We can fix $C_1$ by using the normalization condition $\int_0^L \hat{P}_{\rm st}(x)dx=1$,
\begin{equation}
   \hspace{-2cm} N=\int_0^{L}d x\int_{ x}^{ x+L}d\phi \exp\left[a x+b\cos\left(\frac{2\pi  x}{L}\right)\right]\exp\left[-\left(a\phi+b\cos\left(\frac{2\pi  x}{L}\right)\right)\right]
\end{equation}
Making a variable change $\psi=\phi - x$, we get,
\begin{eqnarray}
   \hspace{-2cm} N & = \int_0^{L}d x\int_{0}^{L}d\psi \exp\left[a x+b\cos\left(\frac{2\pi  x}{L}\right)\right]\exp\left[-\left\{a(\psi+ x)+b\cos\left(\frac{2\pi ( x+\phi)}{L}\right)\right\}\right]\nonumber\\
   \hspace{-2cm} & = \int_0^{L}d x\int_{0}^{L}d\psi \exp\left[-a\psi+2b\sin\left(\frac{2\pi}{L}\psi/2\right)\sin\left(\frac{2\pi}{L}\left( x+\psi/2\right)\right)\right]\nonumber\\
   \hspace{-2cm} & = \frac{L^2}{2\pi} \int_0^{2\pi} d\psi' \exp[-a'\psi']I_0[2b\sin(\psi'/2)].
\end{eqnarray}
Making another variable change $y=\frac{(\pi-\psi')}{2}$ for $0<\psi'<\pi$ and $y=\frac{(\psi'-\pi)}{2}$ for $\pi<\psi'<2\pi$, and $a'=\frac{2\pi}{L}a=\frac{L f}{2\pi\gamma D}$, we get
\begin{eqnarray}
    N&=&\frac{2L^2}{\pi} \exp\left[-L a/2\right] \int_0^{\pi/2} dy \cosh(2a'y)I_0(2b\cos(y))\nonumber\\
    &=&L^2 \exp\left[-L a/2\right] I_{ia'}(b)I_{-ia'}(b),\label{eq:a10}
\end{eqnarray}
where in the second line we have used the property
\begin{equation}
	\int_{0}^{\pi/2}\cos(2\mu x)I_{2\nu}(2a\cos x)dx=\frac{\pi}{2}I_{\nu-\mu}(a)I_{\nu+\mu}(a).
\end{equation}
where the $I_{ia}(x)$ is $a$-th order modified Bessel function
of first kind with purely imaginary order and real argument $x$, and equivalentely, $I_{a}(x) $ is the is $a$-th order modified Bessel function of the first kind and real order~\cite{gradshteyn2014table}.
Using Eqs.~(\ref{eq:a6}) and~(\ref{eq:10}) in Eq.~(\ref{eq:a3}) stationary  probability current  reads
\begin{equation}
 \hat{J}_{\rm st}=\frac{D}{N}(1-\exp\left[-L a\right])=\frac{2D}{L^2}\sinh\left(\frac{L v_f}{2 D}\right)\Big|I_{\frac{iLv_f}{2\pi D}}\Big(\frac{v_k L}{2\pi D}\Big)\Big|^{-2}.
\end{equation}
Now using the above  results, we can also derive the exact form of average angular frequency  in steady state,
\begin{eqnarray}
       \langle \dot x \rangle_{\rm st} & =&\left\langle v_f-v_k\sin\left(\frac{2\pi  x}{L}\right) \right\rangle_{\rm st}=L \hat{J}_{\rm st}\\
       &=& \frac{2D}{L}\sinh\left(\frac{L v_f}{2 D}\right)\Big|I_{\frac{iLv_f}{2\pi D}}\Big(\frac{v_k L}{2\pi D}\Big)\Big|^{-2}.
\end{eqnarray}
From the previous definitions, we can derive an exact form for the average power  inputted into the  system, Eq.~(\ref{eq:10}) in the Main Text, copied here for convenience
\begin{equation}
    \langle \dot{W} \rangle = f\langle\dot{ x}\rangle= \frac{2Df}{L}\sinh\left(\frac{L v_f}{2 D}\right)\Big|I_{\frac{iLv_f}{2\pi D}}\Big(\frac{v_k L}{2\pi D}\Big)\Big|^{-2}.
    \label{a13}
\end{equation}

\underline{ Scaling form}. 
In the limit $f\to k$ and $D\to0$, more precisely  when $(k/f)-1 \sim (1/2)(\gamma D/f)^{-2/3}$, we can approximate  Eq.~(\ref{a13}) as follows.  Defining
\begin{equation}
u=\left(\frac{v_fL}{6\pi D}\right)^{2/3}\left(\frac{k-f}{f}\right),\quad w=\left(\frac{v_fL}{6\pi D}\right)^{1/3}\frac{\langle \dot{W}\rangle }{fv_f},
\end{equation}
we get, following~\cite{stratonovich1967topics}
\begin{equation}
    \hspace{-1.5cm} w=\frac{3}{2\pi u}\left[I^2_{-1/3}(u^{3/2})+I^2_{-1/3}(u^{3/2})+I_{-1/3}(u^{3/2})I_{1/3}(u^{3/2})+I^2_{1/3}(u^{3/2})\right]^{-1}.
\end{equation}
  Taking $u=0$, i.e. $f=k$, we obtain
\begin{eqnarray}
w &= &\frac{3}{2\pi}2^{-\frac{2}{3}}\left[\Gamma\left(\frac{2}{3}\right)\right]^2,
\end{eqnarray}
which implies Eq.~(\ref{scaling}) in the Main Text, copied here for convenience:
\begin{eqnarray}
    \langle \dot{W}\rangle  & =&  \frac{3^{4/3} \Gamma(2/3)^2}{(4\pi)^{2/3}}f v_f^{2/3}\left(\frac{D}{L}\right)^{1/3}.
\end{eqnarray}
where $\Gamma(x)$ is the gamma function. 

\section{Large correlation time limit}  
\label{app:b}
In this section we provide additional details about the derivation of the large-$\tau$ approximation of the average power given by Eq.~(\ref{eqn:largetau}).  
The Fokker-Planck equation for the extended $(x,\eta)$ space corresponding to the two-dimensional Markovian representation in Eqs.~(\ref{eom1}-\ref{eom}) is given by
	\begin{equation}
	    \frac{\partial P}{\partial t}=-\frac{\partial }{\partial x}\left[(v_f-v_k\sin\left(2\pi  x/L\right)+\eta)P\right]+\frac{\partial}{\partial \xi}\left[\left(\frac{\eta}{\tau}+\frac{D_0}{\tau^2}\frac{\partial}{\partial \eta}\right)P\right],
	    \label{2dfp}
	\end{equation}
where $P\equiv P(x,\eta,t)$ is the joint probability density for $x_t=x$, and $\eta_t=\eta$ at time $t$.
In the large correlation time limit  the last term of  Eq.~\ref{2dfp} becomes negligible and  hence the stationary solution is approximatively given by
\begin{equation}
 \frac{\partial }{\partial x}\left[(v_f-v_k\sin\left(2\pi  x/L\right)+\eta)P_{\rm st}(x,\eta)\right]\simeq 0.
\end{equation}
Solving for the steady state solution and taking $f(x)=v_f-v_k\sin{\left(\frac{2\pi  x}{L}\right)}$, we get the following large-$\tau$ approximation of the stationary conditional distribution
\begin{equation}
    P_{\rm st}^{\rm LT}(x|\eta)=\frac{N^{-1}_{\rm LT}}{v_f-v_k\sin{\left(\frac{2\pi  x}{L}\right)}+\eta},
    \label{eq:b3}
\end{equation}
where $N^{\rm LT}$ is a normalization constant. Using Bayes theorem and  Eq.~(\ref{eq:b3})  we find that the stationary joint probability density is given by $P_{\rm st}^{\rm LT}(x,\eta)= P_{\rm st}(\eta)  P_{\rm st}^{\rm LT}(x|\eta)$, where $P_{\rm st}(\eta)=\sqrt{\frac{\tau}{2\pi D}}\exp\left(-\frac{\tau \eta^2}{2D}\right) $ is the stationary density of the active Ornstein-Uhlenbeck noise.
To find the marginal stationary distribution $P_{\rm st}^{\rm LT}(x)$, we   integrate over $\eta$ and assume . Using these assumptions, we get
\begin{eqnarray}
    P^{\rm LT}_{\rm st}(x) &= \int_{-\infty}^{\infty}   P_{\rm st}^{\rm LT}(x|\eta)P_{\rm st}(\eta)d\eta\nonumber \\
    &=\int_{-\infty}^{\infty}\frac{N^{-1}_{\rm LT}}{v_f-v_k\sin {\left(\frac{2\pi  x}{L}\right)}+\eta}\sqrt{\frac{\tau}{2\pi D}}\exp\left(-\frac{\tau \eta^2}{2D}\right) d\eta.
\end{eqnarray}
Defining $\eta'=\sqrt{\frac{\tau}{2 D}}\eta$ and $a=v_f-v_k\sin {\left(\frac{2\pi  x}{L}\right)}$ and then making the change of variable  $y=\eta'+a\sqrt{\frac{\tau}{2 D}}$, we get
\begin{eqnarray}
    P_{\rm st}^{\rm LT}(x)&=&N^{-1}_{\rm LT}\sqrt{\frac{\tau}{2\pi D}}\int_{-\infty}^{\infty}dy\frac{1}{y}\exp\left[-\left(y-a\sqrt{\frac{\tau}{2 D}}\right)^2\right]\nonumber\\
    &=&N^{-1}_{\rm LT}\sqrt{\frac{\tau}{2D}}(i\sqrt{\pi})[\exp(-a'^2)-i\exp(-a'^2) \rm{erfi}(a')],
    \label{pstlt}
\end{eqnarray}
where $a'=a\sqrt{\frac{\tau}{2 D}}$ and $\rm{erfi}$ denotes the imaginary error function. Eq.~(\ref{pstlt}) follows from the property
\begin{equation}
    \int_{-\infty}^{\infty} x^n\exp[-(x-\beta)^2]dx=(2i)^{-n}\sqrt{\pi}H_n(i\beta),
\end{equation}
where $H_n(x)$ is the Hermite polynomial of $n$-th order~\cite{gradshteyn2014table}, and $H_{-1}(i x)=\frac{\sqrt{\pi}}{2}[\exp(-x^2)-i\exp(-x^2) \rm{erfi}(x)]$.
In the large $\tau$ limit, we may use the asymptotic expansion of $\rm{erfi}(a')$ around $a'\to\infty$
\begin{equation}
    \rm{erfi}(a')=-i+\frac{\exp(a'^2)}{\sqrt{\pi}}\left[a'^{-1}+\frac{1}{2}a'^{-3}+O(a'^{-5})\right].
    \label{erfi}
\end{equation}
Using the expansion~(\ref{erfi}) up to third order in~(\ref{pstlt}) we obtain
\begin{eqnarray}
    \fl P_{\rm st}^{\rm LT}(x)\simeq N^{-1}_{\rm LT}\sqrt{\frac{\tau}{2D}}\left[\sqrt{\frac{2D}{\tau}}\frac{1}{v_f-v_k\sin{\left(\frac{2\pi  x}{L}\right)}}+\frac{1}{2}\left(\frac{2D}{\tau}\right)^{3/2}\frac{1}{(v_f-v_k\sin {\left(\frac{2\pi  x}{L}\right)})^3}\right].
    \label{ltdensity}
\end{eqnarray}
To compute the average power inputted  into the system within this approximation, we evaluate the normalization constant by integrating  Eq.~(\ref{ltdensity}):
\begin{eqnarray}
   1= \int_0^L P_{\rm LT}(x)dx= N_{\rm LT}^{-1} \frac{[(2v_f^2+v_k^2)D_a+2(v_f^2-v_k^2)^2\tau]}{2L(v_f^2-v_k^2)^{5/2}\tau}. 
\end{eqnarray}
From this result, we find  the large-$\tau$ approximation of the average power given by Eq.~(\ref{eqn:largetau}), copied here for convenience
\begin{eqnarray}
    \langle \dot{W}\rangle &= f \langle\dot{x}\rangle=f L N^{-1}_{\rm LT}\\
                            &= \left(\frac{2f (v_f^2-v_k^2)^{5/2}\tau}{[(2v_f^2+v_k^2) D_a+2(v_f^2-v_k^2)^2\tau]}\right) .
\end{eqnarray}

\section{Calculation of average power within Fox's approximation and UCNA}
\label{app:c}
Here we  discuss Fox's and  UCNA approximations that we use in Sec.~\ref{sec:FoxUCNA} of the Main Text. These approximations 
were successfully applied  to describe the equilibrium  dynamics of a Langevin system subject to colored noise with  small correlation time i.e. $\sqrt{D\tau}< l_0$, where $l_0$ is the characteristic length  of the system. 
The stationary solution of the Fokker-Planck equation within Fox's approximation~[Eq.~(\ref{eqn:Fox})]
can be computed following similar algebra as in~\ref{app:a}:
\begin{equation}
    \fl P^{\rm Fox}_{\rm{st}}( x)=\frac{1}{N_{\rm{Fox}}}\exp\left(-u_{\rm a}( x)\right)\int_{ x}^{ x+L}d x'\exp\left(u_{\rm a}( x')\right)\left[1+\frac{2\pi v_k\tau}{L}\cos\left(\frac{2\pi x'}{L}\right)\right],\label{eq:PFox}
\end{equation}
where
\begin{eqnarray}
 \fl u_{\rm a}( x) & \equiv & \frac{1}{\Da L}\left[-v_f L x+\frac{v_kL^2}{2\pi}\cos\left(\frac{2\pi x}{L}\right)-\frac{v_k^2L\tau}{4}\cos\left(\frac{4\pi x}{L}\right)-v_fv_kL\tau\sin\left(\frac{2\pi x}{L}\right)\right]\nonumber\\ \fl  &-&\log\left[1+\frac{2\pi v_k\tau}{L}\cos\left(\frac{2\pi x}{L}\right)\right].   
\end{eqnarray}
Note that the first term is  equivalent to the potential of the system for the steady state problem in the Markovian regime.
\begin{figure}[t!]
	\centering
	\includegraphics[width=\textwidth]{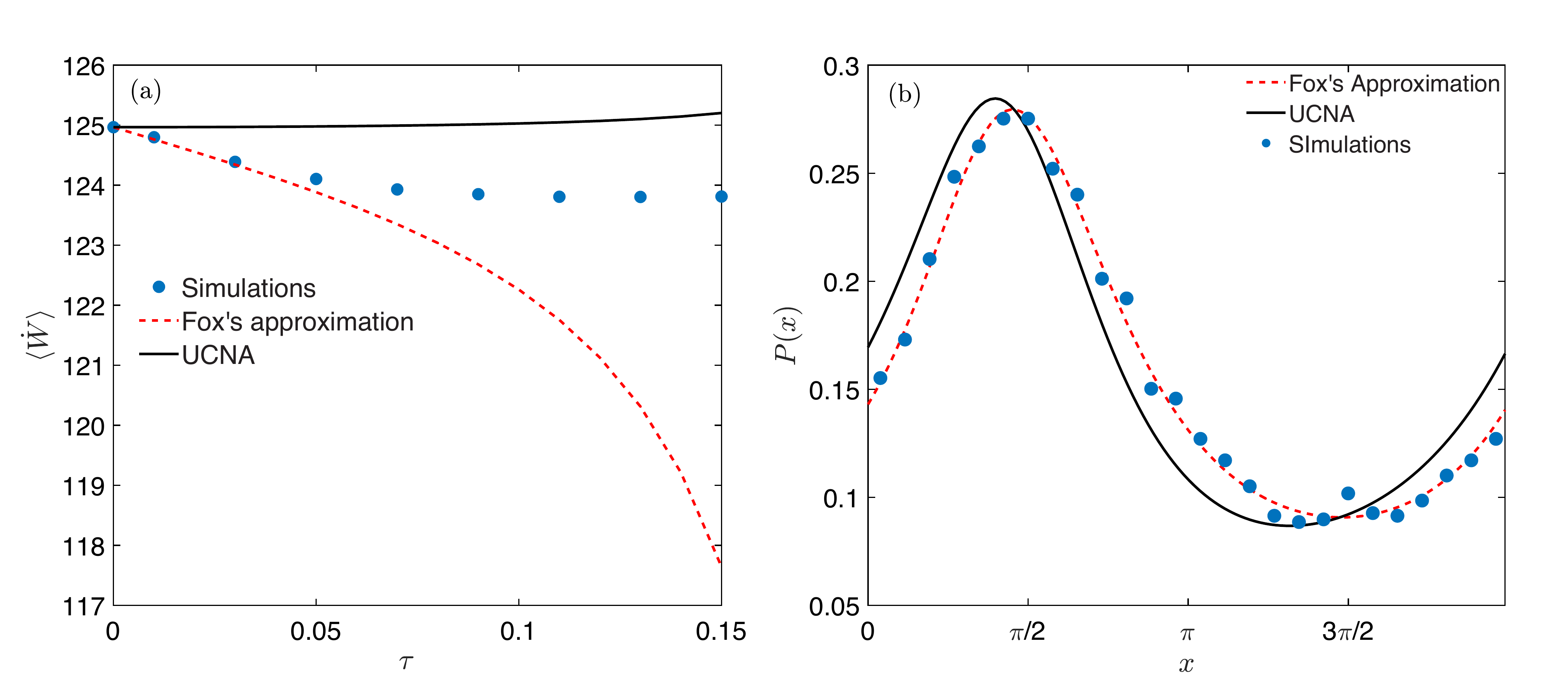}
	\caption{UCNA (solid black line) and Fox's approximation (red dashed line) at small correlation times $\tau< \gamma/(2\pi v_k)$ compared with  results from numerical simulations of Eqs.~(\ref{eom1}-\ref{eom}) (blue circles). (a) Average stationary power as a function of the bacterial bath correlation time. (b)   Stationary distribution of the phase $P_{\rm st}(\phi)$, with $\phi=x$ mod $L$, for  $\tau=0.03$. The solid  and dashed lines are obtained from  numerical integration of  Eqs.~(\ref{eq:PUCNA}) and~(\ref{eq:PFox}) respectively. 
	Parameters: $f=12,k=6,L=2\pi,\gamma=1,D_a=1,\Delta t=10^{-3}$. Statistics for both the panels are computed from $10^{5}$ trajectories.}
	\label{fig:8}
\end{figure}
Similarly, for UCNA \cite{PhysRevA.35.4464}, the stationary distribution of the  Fokker-Planck equation~(\ref{Eqn:ucna})  is given by
\begin{equation}
    \fl P^{\rm UCNA}_{\rm st}( x)=\frac{1}{N_{\rm{UCNA}}}\exp\left(-u_{\rm a}( x)\right)\int_{ x}^{ x+L}d x'\exp\left(u_{\rm a}( x')\right)\left[1+\frac{2\pi v_k\tau}{L}\cos\left(\frac{2\pi x'}{L}\right)\right]^2.\label{eq:PUCNA}
\end{equation}
Therefore, the only difference between these two approximations in the stationary distribution  is the extra factor, $\gamma_{\rm eff}(x)/\gamma=\left[1+\frac{2\pi v_k\tau}{L}\cos\left(\frac{2\pi x'}{L}\right)\right]$ see Eq.~(\ref{eq:sdf}), which enters in the effective underdamped dynamics of the system as a space-dependent friction. Note that such term does not appear in   equilibrium,  as it appears as a space-dependent prefactor in the probability current. Figure~\ref{fig:8} shows that UCNA does not  reproduce accurately  the phase probability density even for  small correlation times when Fox's approximation works. 

 The above considerations imply that the key quantity needed to evaluate the average steady state power is the normalization constant, $N_{\rm{Fox}}=\int_0^{L}d x P^{\rm Fox}_{\rm{st}}( x)  $ and $N_{\rm{UCNA}}=\int_0^{L}d x P^{\rm UCNA}_{\rm{st}}( x)$.  Hence the exact expression for  average power inputted into the system in the  steady state according to Fox's approximation and  UCNA, are given by
\begin{eqnarray}
 \langle \dot{W}_{\rm{Fox}} \rangle =\frac{f D L}{N_{\rm{Fox}}}(1-\exp\left[v_fL/ D\right]),\\
  \langle \dot{W}_{\rm{UCNA}} \rangle =\frac{f D L}{N_{\rm{UCNA}}}(1-\exp\left[v_fL/ D\right]).  
\end{eqnarray}
The above quantities are further evaluated by numerical integration of the above normalization condition, Fig.~\ref{fig:8}a. It must noted that both $N_{\rm{Fox}}$ and $N_{\rm{UCNA}}$ are dependent on all the system parameters.

\begin{figure}[h!]
	\includegraphics[width=\textwidth]{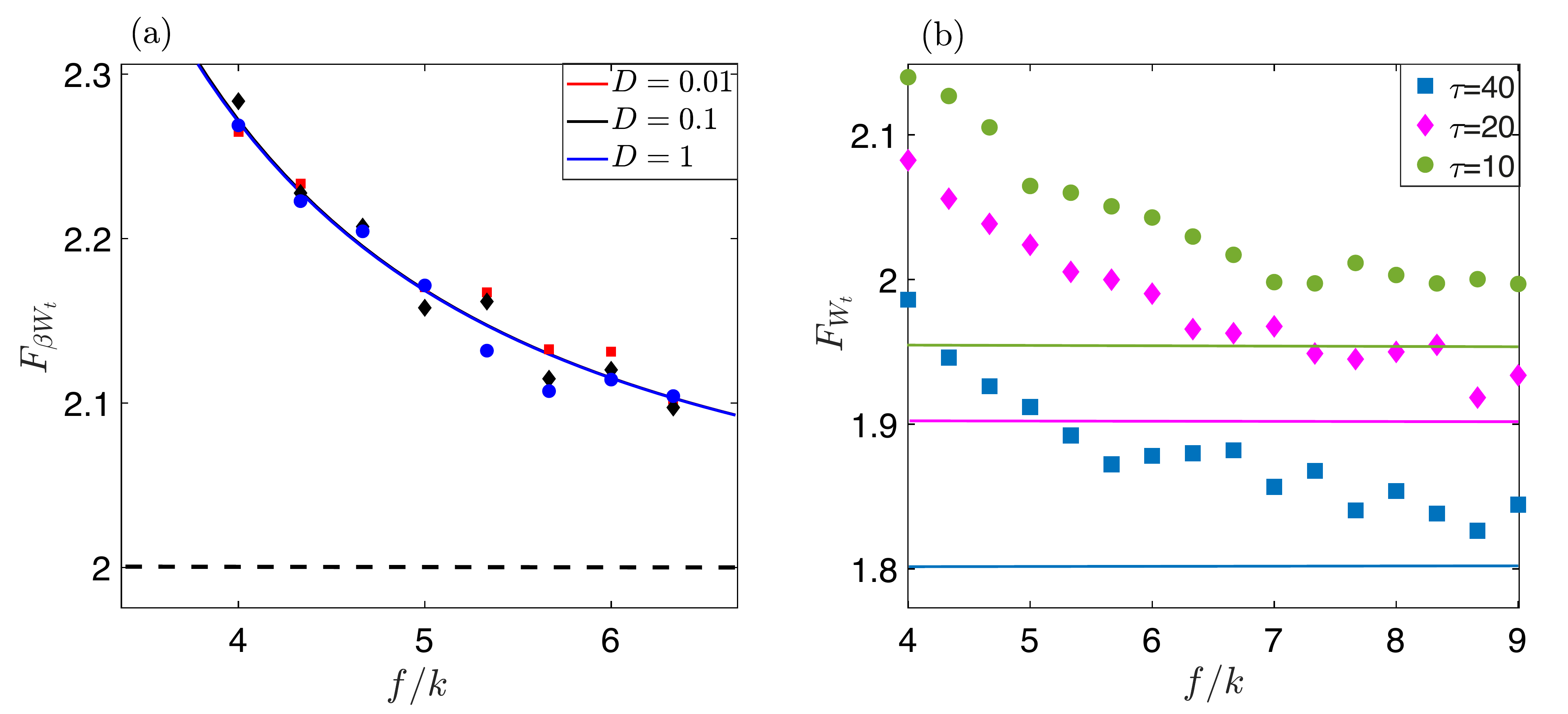} 
	\caption{Uncertainty relations for the work exerted on the oscillator in a thermal (a) and in an active bacterial bath (b). (a) Fano factor $F_{W_t}=\textrm{Var}(W_t)/\langle W_t\rangle$ as a function of the scaled driving torque in the thermal bath limit, for different values of the bath temperature ($T=D/\gamma k_{\rm B}$, with $D$ given in the legend). The dashed line is set to $2k_{\rm B}T$ given by the bound of the thermodynamic uncertainty relation $F_{\beta W_t}\ge 2$~\cite{barato2015thermodynamic,gingrich2016dissipation}. Parameters: $t=800,\gamma=1,L=2\pi,k=6$.   (b) Finite-time Fano factor  for the work done on the oscillator in an active bacterial bath: numerical simulations (symbols) and theoretical bound given by 
 Eq.~(\ref{Eqn:fanoactive})  (solid lines) for different bath correlation times (see legend). Parameters: $t=400,\gamma=1,L=2\pi,k=6,\Da=1,\Delta t=10^{-2}$. Fano factor for both the panels were computed from $10^4$ trajectories.  }
	\label{fig:6}
\end{figure}

\section{Finite-time uncertainty relation for the work exerted in an active bacterial bath}
\label{app:d}
In the thermal bath limit, it can be analytically shown that, at steady state, thermodynamic uncertainty relations saturate in the continuum limit~\cite{barato2015thermodynamic,gingrich2016dissipation} which corresponds to  $f\gg k$ in our model, see Fig. \ref{fig:6}a. In such limit, the  nonlinear periodic force is negligible with respect to the driving torque and hence the dynamics is well approximated by a drift-diffusion process~\cite{risken1996fokker}. Inspired by this observation, we find that the Fano-Factor also reaches a minimum value above the bifurcation point ($f>k$) in the presence of active bath  in the large $f$ limit (see Fig.~\ref{fig:6}b). In this case, it corresponds to a drift-diffusion process subject to Gaussian colored noise
\begin{equation}
	\gamma\dot{x}_t=f+\eta_t,
	\label{drift}
\end{equation}
where $\langle\eta_t\rangle=0$ and $\langle\eta_t\eta_{t'}\rangle=(\gamma^2 D_a/\tau)\exp(-|t-t'|/\tau)$. From the well-known finite-time statistics of the Ornsein Uhlenbeck process $\eta$ we obtain $\langle x_t\rangle= v_f t$ and    $\textrm{Var}[ x_t]=\langle x^2_t\rangle-\langle x_t\rangle^2= 2\Da\left(t-\tau[1-\exp(-t/\tau)]\right)$, which leads to the finite-time work Fano Factor for the drift-diffusion process in the active bath
\begin{eqnarray}
	F^{f\gg k}_{W_t}=\frac{f^2\textrm{Var}[x_t]}{f\langle x_t\rangle}=2\Da\left[1-\frac{1-\exp(-t/\tau)}{t/\tau}\right].
\end{eqnarray}

\section*{References}
\bibliographystyle{naturemag}


\end{document}